\documentclass[11pt]{article}  

\usepackage
{
	amsthm,
        amssymb,
        amsmath,
        graphicx,
        fullpage
}
\pdfoutput=1
\usepackage{amsfonts,latexsym,xspace,makeidx,bm}
\usepackage[usenames]{color}
\usepackage[pagebackref,letterpaper=true,colorlinks=true,pdfpagemode=none,urlcolor=blue,linkcolor=blue,citecolor=red,pdfstartview=FitH]{hyperref}
\usepackage{boxedminipage}

\newcommand{\full}{1}
\newcommand {\rounddown} [1] {{\lfloor {#1} \rfloor}}

\newcommand{\eps}{\varepsilon}

\newcommand{\damp}{\epsilon}

\newcommand {\OT}{\Tilde{\Omega}}
\newcommand{\Gnp}{\mathcal{G}(n,p)}

\newtheorem{theorem}{Theorem}[section]
\newtheorem{lemma}[theorem]{Lemma}

\newtheorem{claim}[theorem]{Claim}
\newtheorem{corollary}[theorem]{Corollary}
\newtheorem{definition}[theorem]{Definition}
\newtheorem{observation}[theorem]{Observation}

\newtheorem{remark}{Remark}[section]

\newtheorem{question}[theorem]{Question}
\definecolor{DSgray}{cmyk}{0,0,0,0.7}

\newcommand{\st}{\mathsf{st}}
\newcommand{\SA}{\text{SA}}

\newcommand{\poly}{\mathrm{poly}}

\newcommand{\F}{\mathbb F}

\usepackage{prettyref}

\newcommand{\savehyperref}[2]{\texorpdfstring{\hyperref[#1]{#2}}{#2}}

\newrefformat{eq}{\savehyperref{#1}{\textup{(\ref*{#1})}}}
\newrefformat{lem}{\savehyperref{#1}{Lemma~\ref*{#1}}}
\newrefformat{def}{\savehyperref{#1}{Definition~\ref*{#1}}}
\newrefformat{thm}{\savehyperref{#1}{Theorem~\ref*{#1}}}
\newrefformat{cor}{\savehyperref{#1}{Corollary~\ref*{#1}}}
\newrefformat{corr}{\savehyperref{#1}{Corollary~\ref*{#1}}}
\newrefformat{cha}{\savehyperref{#1}{Chapter~\ref*{#1}}}
\newrefformat{sec}{\savehyperref{#1}{Section~\ref*{#1}}}
\newrefformat{app}{\savehyperref{#1}{Appendix~\ref*{#1}}}
\newrefformat{tab}{\savehyperref{#1}{Table~\ref*{#1}}}
\newrefformat{fig}{\savehyperref{#1}{Figure~\ref*{#1}}}
\newrefformat{hyp}{\savehyperref{#1}{Hypothesis~\ref*{#1}}}
\newrefformat{alg}{\savehyperref{#1}{Algorithm~\ref*{#1}}}
\newrefformat{item}{\savehyperref{#1}{Item~\ref*{#1}}}
\newrefformat{step}{\savehyperref{#1}{step~\ref*{#1}}}
\newrefformat{conj}{\savehyperref{#1}{Conjecture~\ref*{#1}}}
\newrefformat{fact}{\savehyperref{#1}{Fact~\ref*{#1}}}
\newrefformat{prop}{\savehyperref{#1}{Proposition~\ref*{#1}}}
\newrefformat{claim}{\savehyperref{#1}{Claim~\ref*{#1}}}
\newrefformat{ob}{\savehyperref{#1}{Observation~\ref*{#1}}}

\let\pref=\prettyref

\newcommand{\Sref}[1]{\hyperref[#1]{Section \ref*{#1}}}

\newcommand{\bV}{{\boldsymbol{V}}}
\newcommand{\bU}{{\boldsymbol{U}}}

\newcommand{\E}{\mathop{\bf E\/}}

\renewcommand{\phi}{\varphi}

\newcommand{\norm}[1]{\left\lVert #1 \right\rVert}

\newcommand{\dks}{{\sf Densest $k$-subgraph}}
\newcommand{\kcsp}{{\sf Max $K$-CSP}}

\newcommand{\agr}{{\rm agr}}

\title{Polynomial integrality gaps for strong SDP relaxations of \dks}
\author{Aditya Bhaskara \and Moses Charikar \and Aravindan Vijayaraghavan}
\author{Aditya Bhaskara\thanks{Department of Computer Science, Princeton University, and Center for Computational Intractability. Supported by Supported in part by NSF under CCF 0832797 and AF 0916218.  Email: \texttt{bhaskara@cs.princeton.edu}.} \and Moses Charikar\thanks{Department of Computer Science, Princeton University, and Center for Computational Intractability. Supported in part by NSF under CCF 0832797 and AF 0916218. Email: \texttt{moses@cs.princeton.edu}. } \and Venkatesan Guruswami \thanks{Computer Science Department, Carnegie Mellon University.  Research supported in part by a Packard Fellowship and NSF CCF 1115525. Email: \texttt{guruswami@cmu.edu}.} \and Aravindan Vijayaraghavan\thanks{Department of Computer Science, Princeton University, and Center for Computational Intractability. Supported in part by NSF under CCF 0832797 and AF 0916218. Email: \texttt{aravindv@cs.princeton.edu}. } \and Yuan Zhou \thanks{Computer Science Department, Carnegie Mellon University.  Supported in part by US-Israel BSF grant 2008293, MSR-CMU Center for Computational Thinking, and NSF CCF 1115525.  Email: \texttt{yuanzhou@cs.cmu.edu}.}}
\date{}

\begin{document}

\maketitle

\thispagestyle{empty}
\setcounter{page}{0}

\begin{abstract}
The \dks~ problem (i.e. find a size $k$ subgraph with maximum number of edges),
is one of the notorious problems in approximation algorithms. There is a significant gap between known upper and lower bounds for \dks: the current best
algorithm gives an $\approx O(n^{1/4})$ approximation, while even showing a small constant factor hardness requires significantly stronger assumptions than $\mathrm{P} \neq \mathrm{NP}$. In addition to interest in designing better algorithms, a number of recent results have exploited the conjectured hardness of \dks~ and its variants. Thus, understanding the approximability of \dks~ is an important challenge.

In this work, we give evidence for the hardness of approximating \dks~within polynomial factors. Specifically, we expose the limitations of strong semidefinite programs from SDP hierarchies in solving \dks. Our results include:
\begin{itemize}
\item  A lower bound of $\Omega \big( n^{1/4} / \log^3 n \big)$ on the integrality gap for $\Omega(\log n / \log\log n)$ rounds of the Sherali-Adams relaxation for \dks. This also holds for the relaxation obtained from Sherali-Adams with an added SDP constraint. Our gap instances are in fact Erd\"os-Renyi random graphs.
\item For every $\epsilon > 0$, a lower bound of $n^{2/53-\epsilon}$ on the integrality gap of $n^{\Omega(\epsilon)}$ rounds of the Lasserre SDP relaxation for \dks, and an $n^{\Omega_\epsilon(1)}$ gap for $n^{1-\epsilon}$ rounds. Our construction proceeds via a reduction from random instances of a certain Max-CSP over large domains.
\end{itemize}
In the absence of inapproximability results for \dks, our results show that beating a factor of $n^{\Omega(1)}$ is a barrier for even the most powerful SDPs, and in fact even beating the best known $n^{1/4}$ factor is a barrier for current techniques.

Our results indicate that approximating \dks~ within a polynomial factor might be a harder problem than Unique Games or Small Set Expansion, since these problems were recently shown to be solvable using $n^{\epsilon^{\Omega(1)}}$ rounds of the Lasserre hierarchy, where $\epsilon$ is the completeness parameter in Unique Games and Small Set Expansion.

\end{abstract}

\newpage

\section{Introduction}

The densest $k$-subgraph problem takes as input a graph $G(V,E)$ on $n$ vertices and a parameter $k$, and asks for a subgraph of $G$ on at most $k$ vertices having the maximum number of edges.
%%MOSES: move this defn to Section 3 beginning:
%We denote by density, the average degree of the induced subgraph.
While it is a fundamental graph optimization problem and arises in several applications (community detection in social networks, identifying protein families and molecular complexes in protein-protein interaction networks, etc), there is a huge gap between the best approximation algorithm and the known inapproximability results. The current best approximation algorithm due to \cite{BCCFV} gives $O(n^{1/4 + \eps})$-factor approximation algorithm which runs in time $n^{O(1/\eps)}$ for any constant $\eps>0$.
%\cite{BCCV} gives a combinatorial algorithm that achieves this, as well as one that
%rounds a linear programming relaxation which is weaker than $O(1/\eps)$ rounds of the Sherali-Adams relaxation.
On the inapproximability side, \cite{F02} initially showed a small constant factor inapproximability for \dks~using the random 3-SAT assumption. \cite{khot} used quasi-random PCPs to rule out a PTAS. More recently, \cite{RS,AMM} used more non-standard assumptions to rule out any constant factor approximation algorithms.

While only constant factor approximations have been ruled out, it is commonly believed that \dks~is much harder to approximate even on average (for a natural distribution on hard instances). Recently, average-case hardness assumptions based on the hardness of ``planted'' versions of \dks~were used for public key cryptography \cite{ABW} and in showing that financial derivates can be fraudulently priced without detection \cite{ABBG}.
Given the interest in \dks~from both the algorithms and the complexity point of view,
developing a better understanding of the problem is an important challenge for the field.

In this work, we study lift-and-project relaxations for \dks.
Lift-and-project methods are systematic iterative procedures to obtain sequences of increasingly stronger
mathematical programming relaxations for an integer optimization problem (e.g. Lov\'asz-Schrijver \cite{LS},
Sherali-Adams \cite{SA} and Lasserre \cite{Lasserre}. See the survey by Laurent \cite{Laurent} for a comparison).
Typically, the relaxation obtained after $r$ levels of these strengthenings can be solved in $n^{O(r)}$ time. A number of recent papers have studied the strength and limitations of such relaxations as a basis for designing
approximation algorithms for various problems
 \cite{ABLT,CMM09,Chlamtac,CS,dVKM,GMPT,GMT,KS,RS2,Schoenebeck,STT,STT2,Tul}
 (see the recent survey by Chlamtac and Tulsiani \cite{CT}).
In most cases of approximation algorithms that use strengthened LP and SDP relaxations,  such relaxations can be
obtained from a few levels of such lift-and-project procedures.
In fact, the $O(n^{1/4 + \eps})$ approximation algorithm of \cite{BCCFV} for \dks~
uses a linear programming relaxation which is weaker than that obtained from $O(1/\eps)$ levels of the Sherali-Adams
hierarchy.\footnote{\cite{BCCFV} also gives a purely combinatorial algorithm that does not use a linear program.}
\cite{BCCFV} also show that the integrality gap becomes $O(n^{1/4-\eps})$ after $n^{O(\eps)}$ levels of the Sherali Adams LP hierarchy. %(see \pref{sec:conclusion} for more details).

\paragraph{Our results.} In this paper, we first study lift-and-project relaxations for \dks~ obtained from the Sherali-Adams hierarchy. We show that even $\Omega\big( \frac{\log n}{\log\log n} \big)$ levels of the Sherali-Adams relaxation have an integrality gap of $\OT(n^{1/4})$. Then, we turn to the Lasserre hierarchy for \dks. We show an integrality gap of polynomial ratio ($n^{\epsilon}$, for small enough constant $\epsilon$) for almost linear ($n^{1-O(\epsilon)}$) levels of the Lasserre relaxation. If we only aim at an integrality gap for polynomial ($n^{\epsilon}$) levels of the Lasserre relaxation, the ratio of the gap can be as large as $n^{2/53 - O(\epsilon)}$.

Our gap instances are actually (Erd\"os - Renyi) random graph instances $\mathcal{G}(n,p)$ and random bipartite graphs under a special distribution  -- hence, we show that natural distributions of instances are integrality gap instances with high probability.

We note that prior results exhibiting gap instances for lift-and-project relaxations do so for problems that are {\em already known} to be hard to approximate under some suitable assumption; based on this hardness result, one would {\em expect} lift-and-project relaxations to have an integrality gap that matches the inapproximability factor.

Our gap constructions for \dks~in this paper are a rare exception to this trend, as the integrality gaps we show are substantially stronger than the (very weak) hardness bounds known for the problem.
In fact, we are only aware of the following examples where a polynomial-round Lasserre integrality gap stronger than the corresponding NP-hardness result is known : Max $K$-CSP, $K$-coloring~\cite{Tul}, Balanced Separator and Uniform Sparsest Cut \cite{GSZ11}. In the first two cases, NP-hardness results that are not that far from the gaps are known~\cite{Samorodnitsky-Trevisan,Khot-coloring} and for Max $K$-CSP a matching Unique-Games hardness is also known~\cite{Samorodnitsky-Trevisan09}. For the other two problems, constant factor integrality gaps were shown for linear number of rounds of Lasserre hierarchy \cite{GSZ11}. Again, while these problems are not known to be APX-hard, under the conjectured intractability of Small Set Expansion, they are known to be hard to approximate within any constant factor.

In the absence of inapproximability results for \dks, our results show that beating a factor of $n^{\Omega(1)}$ is a barrier for even the most powerful SDPs, and in fact even beating the best known $n^{1/4}$ factor is a barrier for current techniques. These results are perhaps indicative of the hardness of approximating  \dks~ within $n^{\Omega(1)}$ factors.

\paragraph{Relation to the Small Set Expansion and Unique Games.} A problem related to \dks~ is the Small Set Expansion (SSE) problem, which has received a lot of recent attention due to strong connections to the Unique Games conjecture \cite{RS}. One way to state the SSE conjecture \cite{RS} (which is known to imply the Unique Games conjecture) is as follows: for all $\epsilon > 0$, there exists $\delta, D$ (think of $D$ as a constant), such that the following problem is not polynomial-time solvable:
\begin{definition}[The Gap-SSE problem]
Given a $D$-regular instance $G(V,E)$ with $k=\delta n$, the Gap-SSE problem is to distinguish between the following two cases.
\begin{itemize}
\item \textbf{Yes case.} There exists a subgraph of $k$ vertices with average degree at least $(1-\eps)D$.
\item \textbf{No case.} All subgraphs of $k$ vertices have average degree at most $\eps D$.
\end{itemize}
\end{definition}
Clearly, \dks~ is hard to approximate within any constant factor, assuming the Small Set Expansion conjecture. On the other hand, 
our results indicate that approximating \dks~ even within a polynomial factor may be a harder problem than Unique Games or Small Set Expansion, because these problems were recently shown to be solvable using $n^{\epsilon^{\Omega(1)}}$ rounds of the Lasserre hierarchy, where $\epsilon$ is the completeness parameter in Unique Games and Small Set Expansion \cite{BRS,GS}.

\section{Preliminaries}
\subsection{Notation.}\label{sec:notation}
We introduce some notation which will be used throughout the paper. $G = (V,E)$ refers to a graph which is an instance of the \dks~ problem on $n$ vertices, and $k$ refers to the size of the subgraph we are required to output. For an induced subgraph $H \subseteq G$, we denote by $d(H)$ the average degree (or {\em density} of $H$).  For a vertex $v$ in subgraph $H$, we will denote by $\Gamma_H(v)$ the set of neighbors of $v$ in $H$ (the suffix will be dropped when $H=G$).

The phrase ``with high probability'' will mean: with probability $1-\frac{1}{p(n)}$, for any polynomial $p(n)$.  %It will be clear from the context that there are constants which depend on the degree of $p$.

\subsection{The relaxation hierarchies for \dks.}
We will be concerned with the SDP relaxations derived from the Sherali-Adams and Lasserre hierarchies for \dks. As in other lift-and-project schemes, a feasible solution to $r$ levels of these hierarchies satisfies the condition that for any set of $r$ vertices, it defines a valid distribution over integral solutions for these vertices -- in particular, the integrality gap becomes $1$ after $n$ levels. Further, the relaxations given by $r$ levels of the Sherali-Adams and Lasserre hierarchies can be solved in $n^{O(r)}$ time. We are interested in the integrality gap of $r$ levels of these relaxations for \dks. Refer to \cite{CT} for a more comprehensive comparison of these relaxation hierarchies.

\subsubsection{The Sherali-Adams LP hierarchy.} \label{sec:lp}

The Sherali-Adams hierarchy starts with a simple LP relaxation of a $\{0,1\}$ integer program, and obtains a sequence of successively tighter relaxations with more levels. 
The natural LP relaxation for \dks~ (LP1 in \pref{fig:lps}) \cite{SW,FS} has variables $\{x_{i}\}$ to denote if vertex $i$ belongs to the solution, and edge variables $\{x_{ij}\}_{\substack{(i,j) \in E(G)}}$ to denote if both $i,j$ are in the subgraph. 
\ifnum\full=1
This LP has an integrality gap of $\Omega(\frac{n}{k})$ (\cite{FKP,FS}). 
\fi

\begin{figure}[htbp]
\begin{center}
\begin{tabular}{c|c}
\hline
{\bf Natural LP (LP1):} & {\bf Min. degree LP (LP2):}\\
\hline
\parbox{6cm}{\begin{align*}
 \max & \sum_{(i,j)\in E(G)} x_{ij} \notag \\
 \text{s.t.}~ & \sum_{i\in V} x_i \leq k  \\
 \forall i,j \in V,\quad &  0 \le x_{ij}\le x_i \le 1  \notag
\end{align*}}&
\parbox{6cm}{\begin{align*}
&\max \quad d\\
 \text{s.t.} & \sum_{i\in V} x_i\leq k, \qquad \text{and}\\
 \exists \{x_{ij}\mid i,j\in V\} &\text{ s.t.}\\
 \forall i\in V \quad & \sum_{j\in\Gamma(i)}x_{ij}\geq dx_i\\
 \forall i,j\in V \quad & x_{ij}=x_{ji}\\
 \forall i,j\in V \quad & 0\leq x_{ij}\leq x_i\leq 1
\end{align*}}\\
\hline
\end{tabular}
\end{center}
\caption{Two Linear Programming relaxations for \dks}\label{fig:lps}
\end{figure}
For our integrality gaps, we will in fact start with a stronger basic (first-level) linear program (LP2 in \pref{fig:lps}) which is equivalent upto a factor of $2$ (see \cite{BCCFV}). Intuitively, it tries to find a $k$-subgraph $H$ where the {\em minimum degree} $d_H$ is maximized. An LP hierarchy obtained from this min. degree LP (LP2) was in fact used by \cite{BCCFV} to obtain their approximation algorithm.
\footnote{While the program as stated is not linear, we guess the degree $d$ and consider the feasibility linear program that is obtained.}
%\anote{Maybe we could state that the reason we consider this stronger LP is because it was useful in our algorithm for dks}.

Let us consider strengthening this LP by considering $r$ levels of the Sherali-Adams hierarchy ($SA_r$, shown in \pref{fig:dks-sa}). In the {\em lifted} LP, the variable $x_S$ is supposed to capture whether every vertex in $S$ belongs to the chosen $k$-subgraph (i.e., $x_S = \prod_{i \in S} x_i$).
%Further for any set $S$ of $\le r$ vertices, the variables define a {\em local distribution}, i.e., a distribution over integer solutions for $S$ (this is enforced using the inclusion-exclusion constraints below).
Further if we take two sets $S, S'$ of $\le r$ vertices, the local distributions induced by a feasible solution (using the inclusion-exclusion constraints), agree on the variables in the intersection $S \cap S'$.  We follow the notation established in \cite{CT} while defining the hierarchy.

\subsubsection{The mixed hierarchy (Sherali-Adams + SDP).}
The mixed hierarchy (also refered to as \emph{SA+}) imposes an additional SDP constraint on top of the Sherali-Adams LP relaxation. In particular, it asks for the values $x_{ij}$ to come from vector inner products i.e. the matrix $X=(x_{ij})$ is p.s.d. Most known algorithms which proceed by rounding a relaxation obtained from an SDP hierarchy \cite{Chlamtac,CS,BRS} work with this mixed hierarchy \footnote{\cite{GS} is an exception and seems to need a relaxation given by the Lasserre hierarchy.}.%\anote{Maybe we should mention that Venkat's new result with Sinop uses Lasserre?}
\cite{RS2,KS} and \cite{Tul2} considered this hierarchy and obtained integrality gaps for Unique Games and approximation-resistant CSPs.

\begin{figure}[htbp]
\begin{align}
& \max \quad d \nonumber , \text{~s.t.} \\
& \exists \{x_{S}\mid S \subseteq V, |S| \le r\} \text{ s.t. } x_{\emptyset} =1 \text{ and}\notag \\
\forall S,T \subseteq V \text{ s.t } |S|+|T| \le r :& \notag \\
&\sum_{i \in V} \sum_{J \subseteq T} (-1)^{|J|} x_{S \cup J \cup \{i\}} \leq k \sum_{J \subseteq T} (-1)^{|J|} x_{S \cup J} \label{eq:ub}\\
% \exists \{x_{ij}\mid i,j\in V\} &\text{ s.t.}\\
\forall i\in V \quad  & \sum_{j\in\Gamma(i)} \sum_{J \subseteq T} (-1)^{|J|} x_{S \cup J \cup \{i,j\}}\geq d \sum_{J \subseteq T} (-1)^{|J|} x_{S \cup J \cup \{i\}} \label{eq:lb}\\
 & 0 \leq \sum_{J \subseteq T} (-1)^{|J|} x_{S \cup J} \le 1 \label{eq:inclusion-exclusion}
\end{align}
\caption{Sherali-Adams LP relaxation ($r$ levels) for \dks: $\SA_r$}\label{fig:dks-sa}
\end{figure}

\ifnum\full=1
One level of the mixed hierarchy for \dks~ gives the SDP relaxation introduced in \cite{FS,SW}.  
\fi
\cite{BCCFV} show that the mixed hierarchy performs better than log-density based arguments (which are captured by just the LP hierarchy) in a {\em planted} model.\footnote{In particular, the problem of detecting if dense $k$-subgraph is planted in a random graph or not, in the parameter range $D < n^{1/2}$.} It is interesting in this light to obtain integrality gaps for mixed hierarchy.

\subsubsection{The Lasserre hierarchy.}

The Lasserre hierarchy produces a sequence of SDP relaxations which are stronger than the Sherali-Adams and the mixed hierarchies. As in \cite{CT}, the $r$-level Lasserre SDP for \dks~ introduces a vector $\bU_{S}$ for each subset $S \subseteq V$ with $|S| \leq r$ (\pref{fig:dks-lasserre}).
\begin{figure}[htbp]
\begin{align*}
\text{max } &\sum_{(u, v) \in E} \norm{\bU_{\{u, v\}}}^2\\
\text{ such that }& \\
&\langle \bU_{S_1}, \bU_{S_2}\rangle \geq 0 \text{ for all } S_1, S_2\\
&\langle \bU_{S_1}, \bU_{S_2} \rangle = \langle \bU_{S_3}, \bU_{S_4} \rangle  \text{ when }S_1 \cup S_2 = S_3 \cup S_4\\
&\sum_{v \in V} \langle \bU_{\{v\}}, \bU_{S}\rangle  \leq k \norm{\bU_{S}}^2\text{ for all } S\\
&\norm{\bU_{\emptyset}}^2 = 1
\end{align*}
\caption{Lasserre hierarchy ($r$ levels) for \dks}\label{fig:dks-lasserre}
\end{figure}
The intended solution sets $\bU_{S}=\bU_{\emptyset}$ if every vertex in $S$ belongs to the densest $k$-subgraph, and $\bU_{S}=\bf{0}$ otherwise. The vector lengths $\norm{\bU_{S}}^2$ correspond to valid LP values $x_S$ for the Sherali-Adams relaxation presented above.

\begin{remark}
As in \pref{sec:lp}, we can write an SDP
%(min. degree SDP)
 which tries to find the $k$-subgraph of largest induced minimum degree $d$. This can be captured by the SDP constraint (analogous to \pref{eq:lb})
\begin{equation}\label{eq:sdp-mindeg}
\forall S, \forall u \in V, \quad \sum_{v \in \Gamma(u)} \langle \bU_{\{u,v\}}, \bU_S \rangle \ge d \cdot \langle \bU_{\{u\}}, \bU_S \rangle,
\end{equation}
However, we show in \pref{sec:completeness} that our integrality gaps also hold for the Lasserre hierarchy defined by this SDP. We refer to the SDP with constraint \pref{eq:sdp-mindeg} as the \emph{Min degree Lasserre SDP} .
\end{remark}
%\anote{There is a difference in the program we start out with for the two hierarchies - the feasibility LP used in SA relaxation maybe slightly stronger. We should resolve this.}

\iffalse
\begin{eqnarray*}
\sum_{(u, v) \in E} \norm{\bU_{\{u, v\}}}^2,
\end{eqnarray*}
while subject to
\begin{itemize}
\item $\langle \bU_{S_1}, \bU_{S_2}\rangle \geq 0$ for all $S_1, S_2$,
\item $\langle \bU_{S_1}, \bU_{S_2} \rangle = \langle \bU_{S_3}, \bU_{S_4} \rangle $ when $S_1 \cup S_2 = S_3 \cup S_4$.
\item $\sum_{v \in V} \norm{\bU_{\{v\}}}^2 \leq k$ and $\norm{\bU_{\emptyset}}^2 = 1$.
\end{itemize}
\fi

\section{Integrality Gap for the Sherali-Adams hierarchy}\label{sec:ig}
In what follows $L$ will denote the number of levels of the hierarchy we will consider.
\begin{theorem}\label{thm:main}
Let $L \le \frac{\log n}{10 \log \log n}$.  The integrality gap of $\SA_L$ is at least $\Omega \big( \frac{n^{1/4}}{L \log^2 n} \big)$.
\end{theorem}

To prove \pref{thm:main}, we present instances $G$ where the relaxation has a solution with value $d = \Omega(n^{1/4}/L)$, while the {\em integer optimum}, i.e., the largest density of a $k$-subgraph in $G$ is only $O(\log^2 n)$.  It will be notationally convenient to construct gaps for $L/2$ levels.

\subsection{The instance.}\label{sec:instance}
We in fact give a distribution over instances, and prove that the desired gap holds with high probability.  The instances we consider are $\Gnp$ random graphs with $p = n^{-1/2} \log n$ (thus the expected degree of each vertex is $D=n^{1/2} \log n$). The parameter $k$ is chosen to be $n^{1/2}$.  An easy calculation shows that in any $k$ subgraph, the density (and hence the min-degree) is at most $O(\log^2 n)$ (see full version or \cite{BCCFV,  FKP}). The meat of the argument is thus to show that there exists an LP solution to $\SA_{L/2}$ (Equations~\eqref{eq:ub}-\eqref{eq:inclusion-exclusion}) of value $d=\Omega(n^{1/4}/L)$ even for $L$ of the order $\log n/\log\log n$.

The following are the properties of the distribution $\Gnp$ (with above parameters) we will truly be using [see 
\ifnum\full=1 \pref{sec:proofs} \fi
\ifnum\full<1 full version \fi
for proofs].  Any graph with these properties admits the solution to $\SA_{L/2}$ which we describe.
\begin{enumerate}
\item Every vertex has degree between $(n^{1/2} \log n)/2$ and $2 n^{1/2} \log n$.
\item Any two vertices $i,j$ have at least one common neighbor and has at most $O(\log^2 n)$ common neighbours.
%\item Any two vertices have at least one common neighbour (thus there is a length-two path between any $i,j$).
\end{enumerate}

\subsection{Feasible solution.}

Before formally giving the $x_S$ values, we give intuition as to what they {\em ought to} be.  First, we start out setting $x_i = n^{-1/2}$ (equal for all vertices, since $\sum_i x_i \le k = n^{1/2}$ and no vertex is special).  Next, suppose $S \subset V$ with $i \in S$ and think of $d \approx n^{1/4}$.  Now \eqref{eq:lb} implies that $\sum_{j \in \Gamma(i)} x_{S\cup j} \ge n^{1/4} x_S$.  Further from \eqref{eq:ub}, we obtain $\sum_{j \in V} x_{S\cup j} \le n^{1/2} x_S$.  Thus we conclude that $x_{S\cup j}$ must be roughly $n^{-1/4} x_S$ for $j \in \Gamma(S)$, while for $j \not\in \Gamma(S)$, it should be only $n^{-1/2} x_S$.  Now consider $T \subset V$ which span a tree: we could imagine starting with one vertex and adding vertices one by one (each added vertex is a neighbour of the previous ones), and thus conclude that $x_T$ is roughly $n^{-(|T|+1)/4}$ (since $x_i = n^{-1/2}$ to begin with).  Now let $S$ be an arbitrary set of vertices and consider a tree $T \supseteq S$: by monotonicity (a corollary of \eqref{eq:inclusion-exclusion}), $x_S \ge x_T$, and since this is true for {\em every} such $T$, we need to set $x_S$ to be at least $n^{-(\st(S)+1)/4}$, where $\st(S)$ is the number of vertices (size) in the minimum Steiner tree of $S$.

These, with additional `dampening' factors ($L$-terms), are precisely the values we will set.  More precisely we consider the solution
\begin{equation}\label{eq:assignment}
x_S = n^{-\frac{1}{4} \cdot (\st(S)+1) } \cdot L^{-|S|},
\end{equation}
where $\st(S)$, as above, is the size of the minimum Steiner tree of $S$.
Thus for instance $x_i = n^{-1/2}/L$, while $x_{\{i,j\}} = 1/(n^{3/4} L^2)$ when $(i,j) \in E$ and $1/(n L^2)$ otherwise (the latter is because there is a path of length-2 between any $i,j \in G$ with high probability).

Let us fix $L \le \log n/(10 \log \log n)$.  We now show that the LP solution presented above is feasible for $\SA_{L/2}$ with high probability.  The following lemma is useful in simplifying the analysis: it implies that we need to only consider $T=\emptyset$ while showing that the LP solution satisfies constraints~\eqref{eq:ub} and \eqref{eq:lb}. This is where the 'dampening' factors come into play. 
\ifnum\full<1
Please refer to the full version for the proof.
\fi
\begin{lemma} \label{lem:dominate}
Let $S,T$ be disjoint subsets of $V$ of size at most $t$ and $x_S$ be the solution described above. Then
\[ x_S \ge \sum_{J \subseteq T} (-1)^{|J|} x_{S \cup J}  \ge \frac{x_S}{2} \]
\end{lemma}
\ifnum\full=1
\begin{proof}
One property of the assignment \eqref{eq:assignment} is that $x_{S \cup i} \le x_S/L$ for $i \not\in S$.  Further all the $x_S$ are $\ge 0$, and thus in the sum above, the term corresponding to $J \subseteq T$ contibutes positively when $|J|$ is even and negatively otherwise. Hence,
\begin{align*}
&\ \ \ \  \sum_{J \subseteq T} (-1)^{|J|} x_{S \cup J} \\
& \geq \sum_{\ell =0}^{\ell=\rounddown{\frac{|T|}{2}}} \sum_{\substack{J \subseteq T \\ |J|=2\ell}} \big( x_{S \cup J} - \sum_{i \in T \setminus S} x_{S \cup J \cup \{i\}} \big)\\ 
&\geq \sum_{\ell =0}^{\rounddown{\frac{|T|}{2}}} \sum_{\substack{J \subseteq T \\ |J|=2\ell}} x_{S \cup J} (1 - |T|/L) \\ 
&\geq \sum_{\ell =0}^{\rounddown{\frac{|T|}{2}}} \frac{1}{2} \cdot \sum_{\substack{J \subseteq T \\ |J|=2\ell}} x_{S \cup J} \geq \frac{x_S}{2}  \text{ (since $|T| \le L/2$)}
\end{align*}
A similar proof shows the upper bound, since the $x_{S \cup \{i\}}$ terms for $i \in T$ dominate the contributions of $x_{S \cup J}$ for $|J|>1$. 
\end{proof}
\fi
\begin{corollary}
In checking feasibility, it suffices to check \eqref{eq:ub} and \eqref{eq:lb} with $T = \emptyset$.
\end{corollary}
\ifnum\full=1
\begin{proof}
\pref{lem:dominate} allows us to `remove' the $\sum_{J \subseteq T}$ on both sides of the equations (and set $T=\emptyset$) by losing a factor of 2.  Since we allow constant slack, the claim follows. 
\end{proof}
\fi
We refer to the constraints \eqref{eq:ub} and \eqref{eq:lb} as the {\em size} and the {\em density} constraints respectively, because the former says that we should pick only a $k$-subgraph, and the latter says the minimum degree (density) is at least $d$.  The assignment we described allows us to prove the density constraint easily.

\begin{lemma}(Density Constraint)
The $x_S$ described above satisfy constraints~\eqref{eq:lb}.
\end{lemma}
\begin{proof}
Let $S \subset V$ and $i\in S$.  We need to check that $\sum_{j \in \Gamma(i)} x_{S\cup j} \ge \frac{n^{1/4}}{L} \cdot x_S$.  It is easy to see that for every $j \in \Gamma(i)$, $\st(S\cup j) \le \st(S)+1$, and thus $x_{S \cup j} \ge \frac{n^{-1/4}}{L} \cdot x_S$ (the $L$ term is due to the dependence on $|S|$ in \eqref{eq:assignment}).  Since there are at least $n^{1/2} \log n/2$ terms in the LHS, the inequality follows. 
\end{proof}

\subsection{The Size Constraint and Minimum Steiner trees in $\Gnp$.}
By the above corollary, it suffices to check (noting $k=n^{1/2}$) that
\begin{equation}\label{eq:ub-check}
\sum_{i \in V} x_{S \cup i} \le n^{1/2} x_S \qquad \text{for all } S \subset V, |S|<t.
\end{equation}

We show this by proving that $\st(S\cup i) \ge \st(S)+2$ for {\em most} $i \in V$, in particular we bound the number of exceptions (lemmas below state the precise bounds).  This then implies that \eqref{eq:ub-check} holds.

We start with some basic facts (and notation) about Minimum Steiner trees (minST) of $S (\subset V)$ in $\Gnp$, with our parameters.  We will refer to the vertices in $S$ as the {\em terminals}, and the rest of the vertices in a minST as the {\em non-terminals}.  First, the minST must have all its leaves to be terminals.  Further, since every two vertices in $G$ have a path of length two, we must have $\st(S) \le 2|S|-1$ for all $S$.  This helps us bound the number of {\em tree structures} the minST of $S$ can have.  We define this formally.

Given $S \subset V$, a {\em tree structure} for $S$ is a tree $T$ along with a mapping $g:V(S) \rightarrow V(T)$ which is one-one (not necessarily onto).  The vertices in $T$ without an inverse image in $S$ are called {\em internal vertices} and the rest are also called {\em fixed vertices}.  A tree structure for $S$ is {\em valid} if it is possible to `fill in' the internal vertices with distinct vertices from $V$ such that all the edges in the tree are also present in $G$.  [The relation to Steiner trees is apparent -- the internal vertices are the Steiner vertices].  Given an internal vertex in $T$, the vertices of $G$ which take that {\em position} in some valid `filling in' are called the set of {\em candidates} for that position.

Before we get to the lemmas, we note that the number of tree structures for $S$ of size $\le 2|S|$ is at most $(2|S|)^{2|S|}$ (this is just by a na\"ive bound using the number of trees).  Let us now bound the number of $i \in V$ for which $\st(S \cup i) \le \st(S)+1$.

\begin{lemma}\label{lem:internalbound}
Let $S \subset V$ and $T$ be a tree structure for a min Steiner tree of $S$ (so the leaves of $T$ are elements of $S$).  Then the number of candidates for each of the {\em positions} in $T$ is at most $(\log n)^{2|S|}$.
\end{lemma}
\ifnum\full<1
The proof proceeds by an inductive argument on the size of $S$. The argument uses some structural properties of minimum steiner trees, arising because of the fact that there are paths of length $2$ between any pair of vertices w.h.p. This helps us bound the number of candidates even though minimum steiner trees could have many internal nodes with just one child (degree $2$ internal nodes). Please refer to the full version for the proof.
\fi
\ifnum\full=1
\begin{proof}
The proof is by induction on the size of $S$.  The base case $|S|=1$ is trivial.  Assume the result for all tree structures of sets of size $\le |S|-1$.  Now consider $S$. We may assume that $T$ has at least one non-terminal, as otherwise there is nothing to prove.

First, note that there exists a vertex $u \in T$ which is adjacent to at most one non-leaf vertex in $T$.  This is because deleting all the leaves in $T$ gives a tree (which is not empty as there is at least one non-terminal in $T$), and a leaf in this tree our required $u$.  If $u$ is a terminal, we could remove the leaves attached to $u$ (thus obtaining a subset $S'$ of the terminals), and the remaining tree structure would be a valid min Steiner tree for $S'$. Further, the set of non-terminals is precisely the same, and thus the inductive hypothesis implies the claim for $S$.  Thus suppose $u$ is a non-terminal.

If the degree of $u$ (in $T$) is 2, then $u$ has precisely one leaf attached to it (call it $\ell$).  Consider the tree $T'$ obtained by removing $u, \ell$, and let $S' = S \setminus \ell$.  Now $T'$ is a min Steiner tree for $S'$ (if not, we could consider use this smaller tree for $S'$ along with a path of length $2$ to $\ell$ to obtain a smaller minimum steiner tree for $S$).  If $b$ is the vertex in $T'$ attached to $u$, there are at most $(\log n)^{2|S|-2}$ candidates for $b$, by Induction Hypothesis. For each candidate $b$, the number of candidate $u$ is only $O(\log^2 n)$, and since $\ell$ is a terminal. Thus the number of candidates for $u$ is at most $(\log n)^{2|S|}$.  The rest of the non-terminals in $T$ are also present in $T'$, and this gives the result.

If degree$(u) >2$, then there are at least two leaves attached to $u$, thus the number of candidates for $u$ is only $\log^2 n$.  Consider one candidate $x$ for $u$.  Let $T'$ be the tree obtained by removing all the leaves attached to $u$ (thus $u$ is now a leaf), and $S'$ be $S \cup x$ minus the set of leaves attached to $u$.  Now $T'$ is a min Steiner tree structure for $S'$ (otherwise we can obtain a smaller tree for $S$).  Thus by the inductive hypothesis, the number of candidates for any internal vertex in $T'$ is at most $(\log n)^{2|S|-2}$.  Since there are only $\log^2 n$ of the $x$'s, it follows that the total \#(candidates) for an internal vertex is at most $(\log n)^{2|S|}$.

This completes the proof, by induction. 
\end{proof}
\fi
An easy corollary is the following.
\begin{corollary}\label{corr:equalval}
Let $S \subset V$. There are at most $(2|S|\log n)^{2|S|}$ vertices $i$ such that $\st(S\cup i) = \st(S)$.
\end{corollary}
\begin{proof}
Each such $i$ must be the internal vertex of some min Steiner tree for $S$, and there are at most $(2|S|)^{2|S|}$ tree structures. \pref{lem:internalbound} now implies the claim. 
\end{proof}

\begin{lemma}\label{lem:onemore}
Let $S \subset V$. There are at most $(4|S| \log n)^{4|S|} \times n^{1/2}$ vertices $i$ such that $\st(S \cup i) = \st(S)+1$.
\end{lemma}
\ifnum\full<1
The proof of the lemma proceeds by showing that such vertices $i$ must belong to or be a neighbor of some minimum steiner tree for some $T \subseteq S$, and then invoking \pref{corr:equalval}. We refer the reader to the full version for the proof. 
\fi
\ifnum\full=1
\begin{proof}
Let $i$ be such a vertex.  First, note that if there exists a min Steiner tree for $S \cup i$ with $i$ as a leaf, we are done.  This is because removing $i$ gives a min Steiner tree for $S$, and thus $i$ is a neighbour of an internal vertex in a min Steiner tree for $S$.  Thus by \pref{corr:equalval} there are only $(2|S|\log n)^{2|S|} \times (2n^{1/2} \log n)$ such $i$.

Thus suppose that the min Steiner tree for $S \cup i$ has $i$ as an internal vertex.  We will prove the bound as follows: we consider a tree structure $T$ of size $\st(S)+1$ with leaves being terminals from $S$; then we show that the number of candidates for any fixed position in $T$ is at most $(2|S|\log n)^{2|S|} n^{1/2}$.  This suffices, because the number of choices of tree structures adds an additional factor of $(2|S|)^{2|S|}$.

\begin{figure}[!h]
\begin{center}
\includegraphics[scale=0.50]{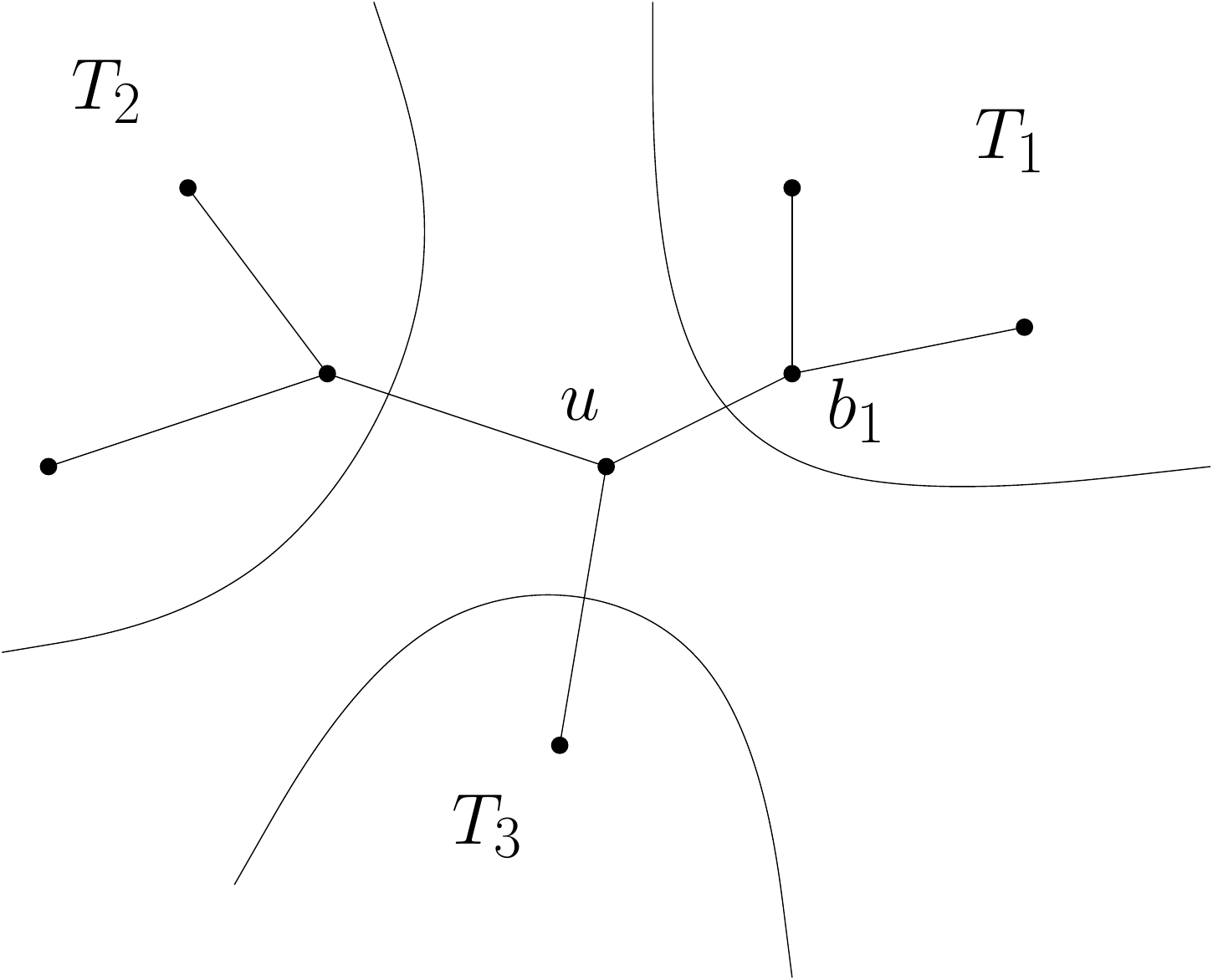}
\end{center}
\caption{An example}
\end{figure}

Let us consider a structure $T$ as above, and a position $u$.  Since $u$ is not a leaf, it has degree at least $2$.  Let the degree be $d$, and let $T_1, \dots, T_d$ be the subtrees of $T$ formed by removing $u$ (see figure ...).  Now if for some $i$, $T_i$ is the min Steiner tree for the terminals in $T_i$, we are done, because then, each candidate for $u$ must be neighbour of an internal vertex in the tree, and by \pref{corr:equalval} there are only $\sqrt{n} \times (2|S|\log n)^{2|S|}$ candidates.  Thus for {\em each} $i$, $T_i$ must have a strictly smaller tree $T_i'$.  Let the vertex in $T_1$ connected to $u$ be called $b_1$.  Now construct a new tree as follows: leave $T_1$ intact, and replace $T_2, \dots, T_d$ by $T_2', \dots, T_d'$; connect $b_1$ to $T_2', \dots, T_d'$ using paths of length $2$.  The number of edges in the new tree is now at most $|T|-d-(d-1)+2(d-1)$. The first term is the original cost, followed by removal of $u$, followed by the decrease by using $T_i'$ as opposed to $T_i$, followed by the cost of adding length-2 paths.

Thus the new tree has cost at most $|T|-1$, and thus it is optimal for $S$! Further, $u$ is adjacent to $b_1$ which is an internal vertex, and thus the number of candidates is bounded by the desired quantity.
\end{proof}
\fi
\paragraph{Putting things together.}  Consider the sum $\sum_{i \in V} x_{S \cup i}$.  \pref{corr:equalval} implies that there are at most $(L \log n)^{L}$ terms which contribute a value $x_S/L$.  \pref{lem:onemore} implies that there are at most $n^{1/2} \cdot (2L \log n)^{2L}$ terms which contribute a value $x_S / (n^{1/4}L)$.  Thus if we pick $(2L \log n)^{2L} < n^{1/4}$, we have the bound that the sum is at most $n^{1/2} x_S$, as desired.

Thus we have verified each of the constraints \eqref{eq:ub}-\eqref{eq:inclusion-exclusion}.  This completes the proof of \pref{thm:main}.

\subsection{Gaps for the mixed hierarchy (SA+).} \label{sec:sdp}

Consider the relaxation $SA_t$ described in \eqref{eq:ub}-\eqref{eq:lb}, along with the constraint: $Z=(x_{ij})_{1\le i,j \le n} \succeq 0$.  The solution considered earlier (Equation~\eqref{eq:assignment}) turns out to also satisfy this PSD condition with high probability.  
\ifnum\full=1
The entries of $Z$ are
\[ Z_{ij} = \begin{cases} n^{-1/2} / L \text{ if } i=j \\n^{-\frac{3}{4}} / L^2 \text{ if } (i,j)\in E(G) \\ n^{-1}/L^2 \text{~ otherwise} \end{cases} \]

Thus we have
\fi
\ifnum\full<1
It is easily seen that \fi
\[ Z = \frac{1}{L} \cdot \Big[ \frac{1}{nL} J + \frac{1}{n^{1/2}} I + \frac{1}{n^{3/4}L} A \Big], \]
where $A$ is the adjacency matrix of $G$.  
\ifnum\full=1
Now $A$ is a $\Gnp$ matrix with $p = n^{-1/2} \log n$.  
Thus the least eigenvalue is at least $-2\sqrt{np(1-p)}$ with high probability (by the Semicircle law).  This is at least $-4n^{1/4} (\log n)^{1/2}$.  
\fi
\ifnum\full<1
Since $A$ is a $\Gnp$ matrix with $p=n^{-1/2}\log n$, its least eigenvalue is at least $-4n^{1/4}(\log n)^{1/2}$ with high probability.
\fi
Thus we have $A + 4n^{1/4}\sqrt{\log n} I \succeq 0$.  Using the fact that $J \succeq 0$, we obtain that $Z \succeq 0$.

%\bnote{right now the $\log n$ terms go the wrong way.. correct this!}
This shows that adding an SDP constraint at the {\em first level} does not give us any additional power -- the relaxation obtained after $\Omega(\frac{\log n}{\log\log n})$ levels also has an integrality gap of $\OT(n^{1/4})$.
\ifnum\full=1
\begin{question}
Does \pref{thm:main} hold for $L \gg \log n$?
\end{question}
\fi
We conjecture that even $L \approx n^{\eps}$ levels does not reduce the integrality gap substantially.  We need a different approach (involving a better argument for bounding the number of trees) to extend the arguments above to this range of $L$.

\iffalse
\newcommand{\vects}[2]{{\bf v}_{#1}^{(#2)}}
\newcommand{\vectb}[1]{{\bf v}_{#1}}

We could appeal to results on eigenvalues of $\Gnp$ or construct a Cholesky decomposition directly as follows: Let $\vects{i}{j}$ represent the $j$ co-ordinate of vector $\vectb{i}$.
\[ \vects{i}{j} = \begin{cases} n^{-1/4 - \damp/2} \text{ if } i=j \\n^{-\frac{1}{2}-\frac{3}{2}\damp} \text{ if } (i,j)\in E(G) \\ n^{-1-\damp} \text{~ otherwise} \end{cases} \]
Using the property that every two vertices $i,j \in V(G)$ has at most $\log^2 n$ common neighbors w.h.p. (\pref{lem:neighbors}), we obtain $Z_{ij}=\vectb{i}\cdot\vectb{j}$.

Since, $\lambda_{\min}(A) \geq - O(\sqrt{D})$, $A + n^{1/4} I \succeq 0$. Since the all ones matrix $J$ is also psd,
\begin{align*}
\frac{1}{n^{1+2\damp}} J+ \frac{1}{n^{3/4+2\damp}}(A+n^{1/4}I) &\succeq 0\\
Z & \succeq 0 \text{ (since $I$ is also p.s.d) }
\end{align*}
\fi

\section{Integrality Gap for the Lasserre hierarchy}

In this section, we show a gap instance with arbitrary large constant ratio for linear-round Lasserre relaxation, and a gap instance with $n^{\eps}$ ratio for $n^{1 - O(\eps)}$-round Lasserre relaxation (\pref{thm:lasserre1}). We also aim at maximizing the ratio of a polynomial-round Lasserre gap instance, getting a ratio of $\Omega(n^{2/53-\eps})$ (\pref{thm:lasserre_final}).

Our construction is based on a variant of Tulsiani's gap instance for \kcsp~ \cite{Tul} -- we extend the parameter range of Tulsiani's instance. Then we convert the \kcsp~ instance to a constraint-variable graph and duplicate the variable vertices, which is our gap instance for \dks. Note that the gap for \kcsp~ problem is indeed a set of random instances. The vector solution from Lasserre gap for \kcsp~ will help us exhibit a good Lasserre vector solution for \dks. We finally use the structure of random instances of \kcsp~ to show the soundness holds with high probability.

Now, let us proceed to the first step, the gap instance for \kcsp.

\subsection{Lasserre Gap for \kcsp~from \cite{Tul}.}

%We follow the framework of Tulsiani \cite{Tul} in transforming Lasserre integrality gap instances through reductions. The starting point of our Lasserre gaps for \dks~ are new Lasserre integrality gaps for the \kcsp~ problem for large values of $K$ (say $n^{\Omega(1)}$). We then exhibit a simple reduction from \kcsp~ problem to \dks~ which helps us to transform lasserre vectors for \kcsp~ problem to lasserre vectors for \dks~ relaxation. Our gap instances for \kcsp~ will, in fact, comprise of \emph{random} instances of our \kcsp~ problem. This allows us to bound from above the integral optimum value of the \dks~ instance that is obtained through the reduction.

We start by defining the \kcsp~ problem.

\begin{definition}
Let $C \subseteq \F_q^K$ be a $q$-ary linear code of block length $K$.
\begin{enumerate}
\item  An instance $\Phi$ of \kcsp$(C)$ is a set of constraints $C_1, C_2, \cdots, C_m$ where each constraint $C_i$ is over a $K$-tuple $T_i = (x_{i_1}, x_{i_2}, \cdots, x_{i_K})$, and is of the form $(x_{i_1} + b^{(i)}_1, x_{i_2} + b^{(i)}_2, \cdots, x_{i_K} + b^{(i)}_K) \in C$ for some $b^{(i)} \in \F_q^K$.

\item A random instance of \kcsp$(C)$ is sampled by choosing each constraint $C_i$ independently, where we sample $K$ variables without replacement from $[n]$ to get $T_i = (x_{i_1}, x_{i_2}, \cdots, x_{i_K})$ and $b^{(i)}$ is chosen from $\F_q^K$ uniformly.
\end{enumerate}
\end{definition}

The following theorem is an extension of the main theorem in \cite{Tul}, showing that polynomial-round Lasserre relaxation cannot refute random \kcsp~ with high probability.

\begin{theorem} \label{thm:Tul09}
If $C$ is the dual code of a distance $2\delta \geq 3$ code (in terms of number of coordinates, not fractional distance), for every $10 \leq K < n^{1/2}$, if $n^{\kappa-1} \leq \eta \leq 1/(10^8 \cdot (\beta K^{2\delta + 0.75})^{1/(\delta - 1)})$ for some $\kappa > 0$, then for large enough $n$, a random instance $\Phi$ of \kcsp$(C)$ over $m = \beta n$  constraints and $n$ variables, with probability $1 - o(1)$, admits a perfect solution for the SDP relaxation obtained by $\eta n/16$ rounds of the Lasserre hierarchy, i.e. there are vectors $\bV_{(S, \alpha)}$ for all $S \subseteq [n]$ with $|S| \leq \eta n/16$ and all $\alpha : S \rightarrow \F_q$, such that
\begin{itemize}
\item the value of the solution is perfect: $\sum_{i = 1}^{m} \sum_{\alpha : T_i \rightarrow \F_q} C_i(\alpha) \norm{\bV_{(T_i, \alpha)}}^2 = m$;

\item $\langle \bV_{(S_1, \alpha_1)}, \bV_{(S_2, \alpha_2)}\rangle \geq 0$ for all $S_1, S_2, \alpha_1, \alpha_2$;

\item $\langle \bV_{(S_1, \alpha_1)}, \bV_{(S_2, \alpha_2)}\rangle = 0$ if $\alpha_1(S_1 \cap S_2) \neq \alpha_2(S_1 \cap S_2)$;

\item $\langle \bV_{(S_1, \alpha_1)}, \bV_{(S_2, \alpha_2)}\rangle = \langle \bV_{(S_3, \alpha_3)}, \bV_{(S_4, \alpha_4)}\rangle$ for all $S_1 \cup S_2 = S_3 \cup S_4$ and $\alpha_1 \circ \alpha_2 = \alpha_3 \circ \alpha_4$;

\item $\norm{\bV_{(\emptyset, \emptyset)}}^2 = 1$ and $\sum_{j \in \F_q} \norm{\bV_{(\{i\}, \{x_i \rightarrow j\})}}^2 = 1$ for all $i \in [n]$.
\end{itemize}
\end{theorem}

Note that \pref{thm:Tul09} extends the original theorem of \cite{Tul} to the regime where $K$ might be superconstant (even $\poly(n)$). The proof of \pref{thm:Tul09} follows the proof in Tulsiani's paper, with the following changes.

\begin{observation}\label{ob:1}
By the first property of the solution given in \pref{thm:Tul09}, we know that for every $i \in [m]$, we have $ \sum_{\alpha : T_i \rightarrow \F_q} C_i(\alpha) \norm{\bV_{T_i, \alpha}}^2 = 1$, and therefore $ \sum_{\alpha : T_i \rightarrow \F_q} C_i(\alpha) \bV_{(T_i, \alpha)} = \bV_{(\emptyset, \emptyset)}$. 
\end{observation}

Recall that Tulsiani showed that, if the constraint-variable graph of a \kcsp$(C)$ instance has very high left-expansion, then the Lasserre SDP admits a perfect solution for it. Formally, the following lemma is (implicitly) shown in \cite{Tul}.
\begin{lemma}[\cite{Tul}]
Given a \kcsp$(C)$ instance, if every set of constraints of cardinality $s \leq r$ involves more than $(K - \delta) s$ variables (where $2\delta$ is the distance of the dual code of $C$), and if $4\delta \leq K$, then there is a perfect solution for the SDP relaxation obtained by $r/16$ rounds of the Lasserre hierarchy.
\end{lemma}
Hence, we only need to prove the following lemma which shows that the constraint-variable graph still has very high left-expansion, even when a constraint might involve superconstant many variables (i.e. the left degree might be superconstant).
%\anote{TODO: We need to give at least a sketch of the theorem proof, also mentioning how the distance of the code comes in.}
\begin{lemma}\label{lem:expansion}
Given $\beta, \eta, K$ as in \pref{thm:Tul09}, with probability $1 - o(1)$, for all $2 \leq s \leq \eta n$, every set of $s$ constraints involves more than $(K - \delta) s$ variables.
\end{lemma}
A similar lemma can be found in \cite{Tul} (Lemma A.1), which only deals with constant $K$. We need a more refined argument for superconstant $K$, which is in \pref{sec:expansionproof}.

\subsection{The Lasserre gap for \dks.}

The gap instance is reduced from the gap instance for \kcsp~in \pref{thm:Tul09}. Let $C$ be the dual code of a $[K, K-t, 2\delta]_q$ code as used in \pref{thm:Tul09}, where $K$ is the block length, $(K-t)$ is the dimension, and $2\delta \geq 3$ is the distance of the code. Such a code has size $|C|=q^t$, and is very sparse for small enough $t$. For $1000 < q$ and $K > q^2$, we let $\beta = (40q^{t+2} \ln q) / K$, and do the following reduction.

Given a \kcsp$(C)$ instance $\Phi$ with $m = \beta n$ constraints and $n$ variables. Let $G_{\Phi} = (L_\Phi, R_\Phi, E_\Phi)$ be the bipartite graph with $m |C|$ left vertices and $nq$ right vertices. For every constraint $C_i$ and every partial assignment to variables in the corresponding tuple $T_i$ which satisfies the constraint $C_i$, we introduce a left vertex.  For every variable $x_i$ and its corresponding assignment, we introduce a right vertex. Formally,
\begin{eqnarray*}
L_\Phi &=& \{(C_i, \alpha) | i \in [m], \alpha : T_i \rightarrow \F_q, C_i(\alpha) = 1\},\\
R_\Phi &=& \{(x_j, \alpha) | j \in [n],  \alpha : \{x_j\} \rightarrow \F_q \} .
\end{eqnarray*}
We connect a left vertex $(C_i, \alpha)$ and right vertex $(x_j, \alpha')$ when $x_j \in T_i$ and $\alpha'$ is consistent with $\alpha$, i.e.
\begin{eqnarray*}
E_\Phi &=& \{\{(C_i, \alpha), (x_j, \alpha')\} | (C_i, \alpha) \in L_\Phi, x_j \in T_i, \alpha'(x_j) = \alpha(x_j) \} .
\end{eqnarray*}

Now we define the final graph $G_\Phi' = (L_\Phi, R_\Phi', E_\Phi')$ in which we want to find a dense $k$-subgraph where $k = 2m$. We take  $\beta$ copies of the right vertices in $R_\Phi$ to get $R_\Phi'$. To get $E_\Phi'$, we connect a left vertex $u \in L_\Phi$ and a right vertex $v \in R_\Phi'$ if $u$ is connected to $v$'s corresponding vertex in $R_\Phi$ in $E_\Phi$. The graph $G_\Phi'$ has $N = m |C| + \beta n q = O(n q^{2t + 2} \ln q / K)$ vertices.

In our analysis of the reduction, we need a $q$-ary linear code $\mathcal{C}$ that has a small constant distance (but no less than $3$), small block length (but more than $q$), and very high dimension. Thus, we instantiate the code $\mathcal{C}$ with Generalized BCH codes given by the following.

\begin{lemma}[Generalized BCH Codes]\label{lem:BCH}
For every prime tower $q$, and integer $2\delta \geq 3$, there are $q$-ary linear codes of block length $K = q^2 - 1$, dimension $(K - 4\delta + 3)$, and distance at least $2\delta$.
\end{lemma}

We include a simple proof of \pref{lem:BCH} as follows.

\begin{proof}
Let $\gamma$ be a primitive element of $\F_{q^2}$. Let $D = 2 \delta$ for notational ease. We construct the following code
\begin{align*}
\tilde{C} = \{(&c_1, c_2, \cdots, c_{q^2 - 1}) \in \F_q^{q^2 - 1} | c(1) = c(\gamma) = c(\gamma^2) = \cdots = c(\gamma^{D-2}) = 0, \\
& \qquad \qquad \qquad \text{where}~ c(X) = c_1 X + c_2 X^2 + c_3 X^3 + \cdots + c_{q^2-1} X^{q^2-1} \} .
\end{align*}

We first show that the distance of $\tilde{C}$ is at least $D$. Since $\tilde{C}$ is a linear code, we only need to show that every non-zero codeword has weight at least $D$.

We show the contrapositive statement : the only codeword of weight at most $D-1$ is $\bm{0}$. For every codeword of weight at most $D-1$, suppose the non-zero entries are in the set $\{c_{i_1}, c_{i_2}, c_{i_3}, \cdots, c_{i_{D-1}}\}$, we have
\begin{align*}
&&&c_{i_1}                     &+&& &c_{i_2}                  &+&&& c_{i_3}                  &+&&& \cdots &&+&& c_{i_{D -1}}                       & = &0\\
&&\gamma^{i_1}& c_{i_1}        &+&& \gamma^{i_2} &c_{i_2}     &+&& \gamma^{i_3}& c_{i_3}     &+&&& \cdots &&+& \gamma^{i_{D-1}}&c_{i_{D -1}}       & =& 0\\
&&\gamma^{2i_1}& c_{i_1}       &+&& \gamma^{2i_2}& c_{i_2}    &+&& \gamma^{2i_3} &c_{i_3}    &+&&& \cdots &&+& \gamma^{2i_{D-1}}&c_{i_{D -1}}      & = &0\\
&&&&&&&& & &  \vdots \\
&&\gamma^{(D-2)i_1}& c_{i_1}   &+&& \gamma^{(D-2)i_2} &c_{i_2} &+&& \gamma^{(D-2)i_3}&c_{i_3} &+&&& \cdots &&+& \gamma^{(D-2)i_{D-1}}& c_{i_{D -1}} & =& 0\\
\end{align*}
Note that the coefficients form a Vandermonde matrix (which has full rank). Therefore we have $c_{i_1} = c_{i_2} = c_{i_3} = \cdots = c_{i_{D-1}} = 0$, i.e. the codeword is $\bm{0}$.

Now we show that the dimension of $\tilde{C}$ is at least $(K - 2D + 3)$. Note that each constraint $c(\gamma^i) = 0 (1 \leq i \leq D-2)$ can be implemented by $2$ linear constraints in $\F_q$ (since $\gamma^i \in \F_{q^2}$), while the constraint $c(1) = 0$ is indeed a linear constraint in $\F_q$. Therefore, we need at most $2(D-2) + 1 = 2D - 3$ linear constraints for $\tilde{C}$, i.e. the dimension of $\tilde{C}$ is at least $(K - 2D + 3)$.

Finally, if the dimension of $\tilde{C}$ is more than $(K - 2D + 3)$, we can take a linear subspace of $\tilde{C}$ of dimension $(K - 2D + 3)$, while the distance of the subspace code is no less than the distance of $\tilde{C}$. 
\end{proof}

\subsection{Analysis.}

We get a family of gap instances $G_\Phi'$ parameterized by $q > 1000$ and $2 \delta \geq 3$ (using \pref{lem:BCH}). We obtain our two main results of this section by picking appropriate parameters for code $C$ as follows. To get lasserre integrality gaps for $N^{1-O(\epsilon)}$ levels , we show the following by setting the distance $2\delta=3$.

\begin{theorem} \label{thm:lasserre1}
For every $1000 < q < N^{\epsilon}$ (where $\epsilon$ is an absolute small constant), there is a gap instance of ratio $\Omega(q)$ for $N/q^{O(1)}$-level Lasserre SDP. The same construction also works for the Min degree Lasserre SDP, when $q = \Omega(\log n)$ and $q < N^{\epsilon}$.
\end{theorem}

We now aim at getting a gap instance of ratio $N^{\epsilon}$ for polynomial-round Lasserre SDP, where $\epsilon$ is maximized.
By setting $q=n^{\gamma}$ for some small constant $\gamma>0$, the distance $2\delta=4$, and optimizing the other parameters, we obtain the following (refer to section~\ref{sec:params} for details)
\begin{theorem}\label{thm:lasserre_final}
For small enough $\kappa > 0$, there is a gap instance of ratio $N^{2/53 - O(\kappa)}$ for the $N^{\kappa}$-round Min degree Lasserre SDP.
\end{theorem}

The two theorems follow because of \pref{thm:Tul09}, \pref{lem:lasserre-dks-completeness}, \pref{lem:completeness-stronger-SDP} (completeness) and \pref{lem:lasserre-dks-soundness} (soundness).
In the completeness case, we will use our $r$-level Lasserre solution for \kcsp~ to show that the Lasserre SDP after $R=r/K$ levels of the hierarchy has value at least $\beta m K$.
In the soundness case, we show that with probability $1 - o(1)$, the graph $G_\Phi'$ does not have any $2m$-subgraph of value more than $17/q$ times the SDP value (\pref{lem:lasserre-dks-soundness}).
Therefore, the graph $G_\Phi'$ is a gap instance of ratio $\Omega(q)$ for $R$-round Lasserre SDP. We proceed by first proving these lemmas.

\subsubsection*{Completeness.} \label{sec:completeness}
\begin{lemma} \label{lem:lasserre-dks-completeness}
If the \kcsp$(C)$ instance $\Phi$ admits a perfect solution for $r$-round Lasserre SDP relaxation, then the $r/K$-round Lasserre SDP relaxation for the \dks~instance $G_\Phi'$ has a solution of value $\beta m K$.
\end{lemma}
\begin{proof}
For any set $S \subseteq L_\Phi \cup R_\Phi'$, suppose the left vertices included in $S$ are
\begin{eqnarray*}
(C_{i_1}, \alpha_1), (C_{i_2}, \alpha_2), \cdots, (C_{i_{r_1}}, \alpha_{r_1}),
\end{eqnarray*}
and the right vertices included in $S$ are
\begin{eqnarray*}
(x_{j_1}, \alpha'_1), (x_{j_2}, \alpha'_2), \cdots, (x_{j_{r_2}}, \alpha'_{r_2}),
\end{eqnarray*}
where $r_1 + r_2 \leq r / K$. Let
\begin{eqnarray*}
S' = T_{i_1} \cup T_{i_2} \cup \cdots \cup T_{i_{r_1}} \cup \{x_{j_1}\} \cup \{x_{j_2}\} \cup \cdots \cup \{x_{j_{r_2}}\}.
\end{eqnarray*}
We have $|S'| \leq K r_1 + r_2 \leq r$. If all the partial assignments $\alpha_i$'s and $\alpha'_i$'s are consistent to each other (i.e. there are not two of them assigning the same variable to different values), we can define
\begin{eqnarray*}
\alpha = \alpha_1 \circ \alpha_2 \circ \cdots \alpha_{r_1} \circ \alpha_1' \circ \alpha_2' \circ \cdots \alpha_{r_2}' ,
\end{eqnarray*}
and let $\bU_{S} = \bV_{(S', \alpha)}$, or we let $\bU_{S} = \bm{0}$.

\begin{observation}\label{ob:2}
For every $S \subseteq L_\Phi \cup R_\Phi'$, we have $\langle \bV_{(\emptyset, \emptyset)}, \bU_S \rangle = \norm{\bU_S}^2$.
\end{observation}
\begin{proof}
If $\bU_S =\bV_{(S', \alpha)}$ for some $S', \alpha$, we have $\langle \bV_{(\emptyset, \emptyset)}, \bU_S \rangle = \langle \bV_{(\emptyset, \emptyset)}, \bV_{(S', \alpha)} \rangle = \norm{\bV_{(S', \alpha)}}^2 = \norm{\bU_S}^2$. If $\bU_{S} = \bm{0}$, we have $\langle \bV_{(\emptyset, \emptyset)}, \bU_S \rangle = \norm{\bU_S}^2 = 0$. 
\end{proof}

We can check that all the Lasserre constraints are satisfied.
\begin{itemize}
\item For two sets $S_1, S_2$, either at least one of the vectors $\bU_{S_1}, \bU_{S_2}$ is $\bm{0}$ (therefore their inner-product is $0$), or $\bU_{S_1} = \bV_{S_1', \alpha_1}, \bU_{S_2} = \bV_{S_2', \alpha_2}$ for some $S_1', S_2', \alpha_1, \alpha_2$ and $\langle \bU_{S_1}, \bU_{S_2} \rangle = \langle \bV_{S_1', \alpha_1}, \bV_{S_2', \alpha_2} \rangle \geq 0$.

\item For any $S_1, S_2, S_3, S_4$ such that $S_1 \cup S_2 = S_3 \cup S_4$, either the set of partial assignments in $S_1 \cup S_2 = S_3 \cup S_4$ are consistent to each other, in which case we have $\bU_{S_1 \cup S_2} = \bU_{S_3 \cup S_4} = \bV_{S, \alpha}$ where $S$ is the union of all the variables included in $S_1 \cup S_2$ and $\alpha$ is the concatenation of the partial assignments in $S_1 \cup S_2$; or we have $\bU_{S_1 \cup S_2} = \bU_{S_3 \cup S_4} = \bm{0}$.

\item For every set $S$, we have
\begin{align*}
& \sum_{v \in L_\Phi \cup R_\Phi'} \langle \bU_{\{v\}}, \bU_S \rangle \\
=&\sum_{(C_i, \alpha) \in L_\Phi} \langle \bU_{\{(C_i, \alpha)\}}, \bU_S \rangle + \sum_{(x_j, \alpha) \in R_\Phi'} \langle \bU_{\{(x_j, \alpha)\}}, \bU_S \rangle\\
=&\sum_{(C_i, \alpha) \in L_\Phi} \langle \bV_{\{(T_i, \alpha)\}}, \bU_S \rangle  + \sum_{(x_j, \alpha) \in R_\Phi'} \langle \bV_{(\{x_j\}, \alpha)}, \bU_S \rangle\\
=& \sum_{i = 1}^{m} \sum_{\alpha : T_i \rightarrow \F_q} C_i(\alpha)\langle \bV_{\{(T_i, \alpha)\}}, \bU_S \rangle   + \beta \sum_{j = 1}^{n} \sum_{\alpha : \{x_j\} \rightarrow \F_q} \langle \bV_{(\{x_j\}, \alpha)}, \bU_S \rangle\\
=& \sum_{i = 1}^{m} \langle V_{(\emptyset, \emptyset)}, U_S \rangle + \beta \sum_{j = 1}^{n} \sum_{\alpha : \{x_j\} \rightarrow \F_q} \langle V_{(\emptyset, \emptyset)}, U_S\rangle\\
=&(m + \beta n) \norm{U_S}^2 = 2m \norm{U_S}^2,
\end{align*}
where the third last equality is because of \pref{ob:1}, and the second last equality is because of  \pref{ob:2}.
\item Finally, we have $\norm{\bU_{\emptyset}}^2 = \norm{\bV_{(\emptyset, \emptyset)}}^2 = 1$.
\end{itemize}

Now, we calculate the value of the solution
\begin{align*}
& \sum_{(u, v) \in E_\Phi'} \norm{\bU_{\{u, v\}}}^2 \\
= & \beta \sum_{(u, v) \in E_\Phi} \norm{\bU_{\{u, v\}}}^2  \\
=&\beta \sum_{i = 1}^{m} \sum_{\alpha : T_i \rightarrow \F_q, C_i(\alpha) = 1} \sum_{x_j \in T_i} \norm{\bU_{\{(C_i, \alpha), (x_j, \alpha_{|\{x_j\}})\}}}^2\\
= &\beta \sum_{i = 1}^{m} \sum_{\alpha : T_i \rightarrow \F_q, C_i(\alpha) = 1} K\norm{\bV_{(T_i, \alpha)}}^2 \\
=& \beta \sum_{i = 1}^{m} K =  \beta m K .
\end{align*} 
\end{proof}

If we add the constraint \pref{eq:sdp-mindeg}, we can still get a good SDP solution for the Min degree Lasserre SDP with high probability, as long as $q$ is superconstant.
\begin{lemma}\label{lem:completeness-stronger-SDP}
For $q = \Omega(\log n)$, with probability $1 - o(1)$, this vector solution also satisfies the added constraint \pref{eq:sdp-mindeg} with $d=\beta K/2$, i.e., for every set $S$, for each vertex $u$, we have
\[ \forall S, \forall u \in V, \quad \sum_{v \in \Gamma(u)} \langle \bU_{\{u,v\}}, \bU_S \rangle \ge \beta K /2 \cdot \langle \bU_{\{u\}}, \bU_S \rangle. \]
\end{lemma}

\begin{proof}
For each left vertex $(C_i, \alpha)$, we have
\begin{align*}
&\sum_{v \in \Gamma((C_i, \alpha))} \langle \bU_{\{(C_i, \alpha),v\}}, \bU_S\rangle\\
=& \beta \sum_{x_j \in T_i} \langle \bU_{\{(C_i, \alpha),(x_j, \alpha_{|\{x_j\}})\}}, \bU_S \rangle \\
=& \beta \sum_{x_j \in T_i} \langle \bU_{\{(C_i, \alpha)\}}, \bU_S \rangle = \beta K \langle \bU_{\{(C_i, \alpha)\}}, \bU_S \rangle  .
\end{align*}

For each right vertex $(x_j, \alpha')$, we have
\begin{align*}
&\sum_{v \in \Gamma((x_j, \alpha'))}  \langle \bU_{\{(x_j, \alpha'),v\}}, \bU_S \rangle \\
=& \sum_{i : T_i \ni x_j} \sum_{\alpha : T_i \rightarrow \F_q, C_i(\alpha) = 1, \alpha(x_j) = \alpha'(x_j)}  \langle \bU_{\{(x_j, \alpha'),{(C_i, \alpha)}\}}, \bU_S \rangle \\
 =& \sum_{i : T_i \ni x_j} \sum_{\alpha : T_i \rightarrow \F_q, C_i(\alpha) = 1, \alpha(x_j) = \alpha'(x_j)}  \langle \bU_{\{{(C_i, \alpha)}\}}, \bU_S \rangle \\
=& \sum_{i : T_i \ni x_j} \sum_{\alpha : T_i \rightarrow \F_q, \alpha(x_j) = \alpha'(x_j)}   \langle \bU_{\{{(C_i, \alpha)}\}}, \bU_S \rangle ,
\end{align*}
where the last equality is because we know that $\bU_{\{(C_i, \alpha)\}} = \bm{0}$ when $C_i(\alpha) \neq 1$. By the property of Lasserre vectors, we know that for each $i \in [m]$,
\begin{align*}
\sum_{\alpha : T_i \rightarrow \F_q, \alpha(x_j) = \alpha'(x_j)}   \bU_{\{{(C_i, \alpha)}\}}  = \bU_{\{{(x_j, \alpha')}\}}, 
\end{align*}
therefore
\begin{align*}
\sum_{v \in \Gamma((x_j, \alpha'))}\langle \bU_{\{(x_j, \alpha'),v\}}, \bU_S \rangle = \sum_{i : T_i \ni x_j} \langle\bU_{\{{(x_j, \alpha')}\}}, \bU_S \rangle .
\end{align*}

For $q = \Omega(\log n)$, the expected number of constraints containing $x_j$ is $\beta K = \Omega((\log n)^{t+2}) = \Omega(\log n)$, by our choice of $\beta$. Therefore, by Chernoff bound and union bound, with probability $1 - o(1)$, for all $x_j$, there are at least $\beta K / 2$ constraints containing $x_j$, and for every $S$, $x_j$ and $\alpha'$, we have
\begin{align*}
&\sum_{v \in \Gamma((x_j, \alpha'))}  \langle \bU_{\{(x_j, \alpha'),v\}}, \bU_S \rangle  \geq \beta K / 2 \cdot \langle \bU_{\{{(x_j, \alpha')}\}}, \bU_S \rangle .
\end{align*} 
\end{proof}

\subsubsection*{Soundness.}
Now, we show that random instances of \kcsp~ give rise to graphs $G'_{\phi}$ whose $2m$-sized subgraphs have density $O(\beta K/q)$. Note that the large alphabet size $q$ allows us to get a much larger gap than we would starting from random AND instances \cite{F02}. This allows us some slack in the size of the subgraphs we need to argue about.

For $C$ the dual of a $[K, K-t, 2\delta]_q$ code, we prove the following soundness lemma.
\begin{lemma}\label{lem:lasserre-dks-soundness}
When $\beta\geq  (40q^{t+2} \ln q) / K$, for a random \kcsp$(C)$ instance $\Phi$, with probability $1 - o(1)$, any subgraph of $G_\Phi'$ obtained by choosing $2m$ left vertices and $2m$ right vertices contains at most $17\beta m K / q$ edges, and therefore any $2m$-subgraph of $G_\Phi'$ contains at most $17\beta m K /q$ edges.
\end{lemma}

Note that $G'_{\phi}$ was constructed by taking $\beta$ copies of the right bipartition and replicating the edges. To prove \pref{lem:lasserre-dks-soundness}, we only need to prove the following lemma.
\begin{lemma}\label{lem:dks-soundness-1}
Suppose that $q > 1000, K>q^2/2, t \leq 10$. When $\beta \geq (40q^{t+2} \ln q) / K$, for a random \kcsp$(C)$ instance $\Phi$, with probability $1 - o(1)$, any subgraph of $G_\Phi$ obtained by choosing $2m$ left vertices and $2n$ right vertices contains at most $ 17m K / q$ edges.
\end{lemma}
\begin{proof}[Proof of \pref{lem:lasserre-dks-soundness} from \pref{lem:dks-soundness-1}]
We only need to prove once there is a $2m \times 2m$ subgraph of $G_\Phi'$ with $t$ edges, there is a $2m \times 2n$ subgraph of $G_\Phi$ with at least $t/\beta$ edges. Fix $2m$ left vertices in $G_\Phi'$, to maximize the number of edges in the subgraph, we need to select the $2m$ right vertices with most edges connected to the chosen $2m$ left vertices. Since any two right vertices $G_\Phi'$ corresponding to the same right vertex in $G_\Phi$ have the same set of neighbors, there is an densest $2m \times 2m$ subgraph $H'$ of $G_\Phi'$ that, for any two such vertices, chooses either both or neither of them. Now we define an subgraph $H$ of $G_\Phi$ that contains the same $2m$ left vertices. It contains a right vertex if any copy of the vertex is contained in $H'$. $H$ contains $2m / \beta = 2n$ vertices, and it is easy to see that there are (at least) $t / \beta$ edges in $H$. 
\end{proof}

We proceed by fixing a set of $2n$ vertices $R$ on the right. \pref{lem:dks-soundness-1} follows from the following claim by a standard union bound over all possible choices of $R$.
\begin{claim} \label{claim:dks-soundness-2}
Recall that $G_\Phi = (L_\Phi, R_\Phi, E_\Phi)$. Suppose that $q > 1000, K > q^2/2, t \leq 10$. Fix a subset $R \subseteq R_\Phi$ (note that $R_\Phi$ is the same for all the instances $\Phi$ of $n$ variables), $|R| = 2n$, the probability (over choice of $\Phi$) that there does not exist a subset $L \subseteq L_\Phi$ of size $2m$ such that the number of edges in the induced subgraph by $L \cup R$ is more than $17m K/q$, is at least $1 -  \exp(-mK/(10q^{t+2}))$.
\end{claim}

\begin{proof} [Proof of \pref{lem:dks-soundness-1} from \pref{claim:dks-soundness-2}]
Since there are only ${qn \choose 2n} \leq \exp(2n(\ln q + 1))$ choices of $R$, by a union bound, with probability at least
\begin{align*}
&1 - \exp(2n(\ln q + 1)) \cdot  \exp(-mK/(10q^{t+2})) \\
&\qquad\qquad = 1 - \exp(2n(\ln q + 1) -\beta n K/(10q^{t+2})) ,
\end{align*}
there is no $2m \times 2n$ subgraph of $G_\Phi$ containing more than $17 mK/q$ edges. The probability becomes $1 - o(1)$ when $\beta = (40q^{t+2} \ln q) / K $. 
\end{proof}

%Now we are going to prove \pref{lem:dks-soundness-2}.
\begin{proof}(Proof of \pref{claim:dks-soundness-2})
First, we show that with high probability, a constraint $C_i$ is ``poorly satisfied''. That is, none of the left vertices corresponding to a constraint $C_i$ has more than $\Omega(K/q)$ neighbors in $R$ -- this number is roughly $1/q$ times the corresponding value in completeness case. We prove this in the following two steps.

\noindent\textbf{Step 1.} Fix a subset $R \subseteq R_\Phi$, $|R| = 2n$, for each variable $x_j$, let $\deg(x_j)$ be the number of vertices in $R$ that corresponding to $x_j$, i.e. let $\deg(x_j) = |R \cap \{(x_j, \alpha) | \alpha : \{x_j\} \rightarrow \F_q\}|$. For a subset of variables $T \subseteq \{x_1, x_2, \cdots, x_n\}$, let $\deg(T) = \sum_{x_j \in T} \deg(x_j)$. We call $T$ \emph{good} if the average degree of variables in $T$ is not more than $4$, i.e. $\deg(T) \leq 4 |T|$.

For a random $T$ with $|T| = K$, note that the expected degree $\E[\deg(T)] = 2K$. Therefore, by Hoeffding's inequalities for sampling without replacement (Theorem 1 and Theorem 4 in \cite{Hoeffding}),
% for sampling without replacement,
we have
\begin{align*}
&\Pr[\text{$T$ is not good}] = \Pr[\deg(T) > 4K] < \exp(-\ln(4/e) \cdot 2K/q) < \exp(-K/(2q)) .
\end{align*}

\noindent\textbf{Step 2.} Again, fix $R \subseteq R_\Phi$, $T \subseteq \{x_1, x_2, \cdots, x_n\}$, for a codeword $\alpha$ on coordinates in $T$, i.e. $\alpha : T \rightarrow \F_q$, let $\agr_T(\alpha, R) = | \{(x_j, \alpha_{|x_j}) | x_j \in T\} \cap R |$. For a constraint $C_i$, say it is \emph{poorly satisfied} if for all $\alpha: T \rightarrow \F_q$ such that $C_i(\alpha) = 1$, we have $ \agr_{T_i}(\alpha, R) \leq 8 K/q $.

Recall that to sample a random constraint $C_i$, we first sample a random $K$-tuple $T_i$, and a random shifting function $b^{(i)}$. Note that for a fixed $\alpha : T \rightarrow \F_q$, and a fixed $T_i$ that is good, when we take a random shifting function $b^{(i)} : T_i \rightarrow \F_q$, we have $\E_{b^{(i)} : T_i \rightarrow \F_q} [\agr_{T_i}(\alpha - b^{(i)}, R)] = \deg(T_i)/q = 4K/q$, therefore, by standard Chernoff bound, for a fixed codeword $\alpha \in C$, the probability that $\alpha$ makes $C_i$ not poorly satisfied is bounded from above by
\begin{align*}
&\Pr_{b^{(i)}}[\agr_{T_i} (\alpha - b^{(i)}, R) > 8K/q]  < \exp(-\ln(4/e) \cdot 4K/q) < \exp(-K/q) .
\end{align*}
Since there are $|C| = q^t \leq q^{10}$ codewords, by a union bound, for $K > q^2/2$ and $q > 1000$, we have
\begin{align*}
&\Pr[\text{$C_i$ is not poorly satisfied} | \text{$T_i$ is good}]  < q^{10} \cdot \exp(-K/q) < \exp(-K/(2q)) .
\end{align*}
In all, we have
\begin{align*}
& \Pr[\text{$C_i$ is poorly satisfied}] \\
\geq & \Pr[\text{$C_i$ is poorly satisfied} | \text{$T_i$ is good}] \cdot \Pr[\text{$T$ is good}]\\
> & (1 - \exp(-K/q)) (1 - \exp(-K/(2q)))\\
> & 1 - \exp(-K/(3q)).
\end{align*}

Now, again, by standard Chernoff bound, we have
\begin{align*}
&\Pr[|\{C_i | \text{$C_i$ is not poorly satisfied}\}|  > m/(q \cdot |C|) ] \\
<& \left(e \cdot |C| \cdot q \cdot \exp(-K/(3q)) \right)^{m/(q \cdot |C|)} \\
\leq& \left(e \cdot q^{10} \cdot q \cdot \exp(-K/(3q)) \right)^{m/(q \cdot |C|)}\\
<& \exp(-K/(10q)) ^{m/(q \cdot |C|)} \\
=& \exp(-mK/(10q^{t+2})).
\end{align*}

By the calculation above we know that with probability at least $1 -  \exp(-mK/(10q^{t+2})) $, there are at most $m/(q \cdot |C|)$ constraints that are not poorly satisfied.

For each left vertex $(C_i, \alpha) \in L_\Phi$, if $C_i$ is poorly satisfied, we know there are at most $8K/q$ edges from $(C_i, \alpha)$ to $R$. If $C_i$ is not poorly satisfied, there are at most $K$ edges to $R_\Phi$ -- this upperbound also applies to $R$.

Therefore, with probability at least $1 -  \exp(-mK/(10q^{t+2}))$, any set of $2m$ left vertices has at most $2m \cdot 8K/q + m /(q \cdot |C|) \cdot |C| \cdot K \leq 17 m K/q$ edges connected to $R$. 
\end{proof}

We now complete the proofs of the main theorems in this section.

\subsubsection*{Proof of \pref{thm:lasserre1}.} \label{sec:lasserre1}

By combining \pref{thm:Tul09}, \pref{lem:lasserre-dks-completeness}, \pref{lem:completeness-stronger-SDP} (completeness), and \pref{lem:lasserre-dks-soundness} (soundness) we see that with probability $1-o(1)$, the graph $G'_{\Phi}$ provides a $\Omega(q)$ integrality for the number of levels $R$ given by

\begin{align*}
 R & = \Omega\left(\frac{n}{K(\beta K^{2\delta + 0.75})^{1/(\delta - 1)}}\right) \\
 &=  \Omega\left(\frac{N}{q^{2t + 2} \ln q \cdot (\beta K^{2\delta + 0.75})^{1/(\delta - 1)}}\right) \\
 & = \Omega\left(\frac{N}{K^{(2\delta - 0.25)/(\delta - 1)}q^{(2t + 2) + (t+2)/(\delta - 1)} \poly\log q}\right).\\
 & \qquad\qquad\qquad \qquad \text{(recall that $\beta = (40q^{t+2} \ln q) / K$)}
\end{align*}

Recall that $K=q^2-1$. By setting $K=q^2-1$ and $2\delta=3$, we verify that the theorem holds.  \qed
%\end{proof}

\subsubsection*{Proof of \pref{thm:lasserre_final}.}\label{sec:params}
Let $q = n^{\gamma}$, since $N = O(n q^{2t+2}\ln q / K) = O(n q^{2t} \ln q)$, ratio of the gap due to \pref{lem:lasserre-dks-completeness} and \pref{lem:lasserre-dks-soundness} is
\begin{align*}
&\Omega(q) = \Omega(N^{\gamma/(1 + 2t \gamma + o(1))}) = \Omega(N^{\gamma/(1 + (8\delta - 6) \gamma + o(1))}) . \qquad \qquad \text{(note that $t = 4\delta - 3$)}
\end{align*}
This means that
\begin{align*}
\epsilon &= \frac{\gamma}{1 + (8\delta - 6) \gamma + o(1)} = \frac{1}{8\delta - 6} - \frac{1 + o(1)}{(1 + (8\delta - 6) \gamma + o(1))(8 \delta - 6)} .
\end{align*}
Note that when $2\delta \geq 3$ is fixed, $\epsilon$ is maximized when $\gamma$ is maximized.

The number of rounds (due to \pref{thm:Tul09} and \pref{lem:lasserre-dks-completeness}) is
\begin{align*}
&R = \Omega\left(\frac{N}{K^{(2\delta - 0.25)/(\delta - 1)}q^{(2t + 2) + (t+2)/(\delta - 1)} \poly\log q}\right)\\
&= \Omega\left(\frac{N}{n^{2\gamma (2\delta - 0.25)/(\delta - 1)}\cdot n^{\gamma ((8\delta - 4) + (4\delta - 1)/(\delta - 1)) + o(1)}}\right)\\
& \quad\qquad\qquad \qquad \text{(by $K = \Theta(q^2)$, $q = n^\gamma$, $t = 4\delta - 3$)}\\
&= \Omega\left(\frac{N}{n^{\gamma(8 \delta + 4 + 6.5/(\delta - 1)) + o(1)}}\right)\\
&= \Omega\left(N^{1 - \frac{\gamma(8 \delta + 4 + 6.5/(\delta - 1)) + o(1)}{1 + o(1) + \gamma(8\delta - 6)}}\right)
\end{align*}

For very small $\kappa > 0$, to get a gap instance for $N^{\Omega(\kappa)}$-round Lasserre, we need
\begin{align*}
&1 - \frac{\gamma(8 \delta + 4 + 6.5/(\delta - 1)) + o(1)}{1 + o(1) + \gamma(8\delta - 6)} \geq \Omega(\kappa)\\
\Rightarrow & 1 + o(1) + \gamma(8\delta - 6)  - (\gamma(8 \delta + 4 + 6.5/(\delta - 1)) + o(1)) \geq \Omega(\kappa) \\
\Rightarrow & 1 - \gamma(10 - 6.5(\delta - 1)) \geq \Omega(\kappa) \\
\Rightarrow & \gamma \leq \frac{1 - \Omega(\kappa)}{10 - 6.5(\delta - 1)}
\end{align*}

Let $\gamma = \frac{1 - O(\kappa)}{10 + 6.5/(\delta - 1)}$, we have
\begin{align*}
& \epsilon = \frac{1}{8\delta - 6} - \frac{1 + o(1)}{(1 + (8\delta - 6) \gamma + o(1))(8 \delta - 6)} \\
= &\frac{1}{8\delta - 6} - \frac{1 + o(1)}{(1 + (8\delta - 6)\frac{1 - O(\kappa)}{10 + 6.5/(\delta - 1)} + o(1))(8 \delta - 6)}\\
= &\frac{1}{8\delta - 6} - \frac{1}{(1 + \frac{(8\delta - 6)}{10 + 6.5/(\delta - 1)})(8 \delta - 6)} - O(\kappa) .
\end{align*}
When $2\delta = 4$, we get the maximized value $\epsilon = 2/53 - O(\kappa)$.  \qed
%\end{proof}

\subsection{Expansion for random \kcsp~ instances.} \label{sec:expansionproof}

In this section, we prove \pref{lem:expansion}, restated as follows.

\vspace{2ex}
\noindent \textbf{\pref{lem:expansion} (restated).}
{\it Given $\beta, \eta, K$ as in \pref{thm:Tul09}, with probability $1 - o(1)$, for all $2 \leq s \leq \eta n$, every set of $s$ constraints involves more than $(K - \delta) s$ variables.}

\begin{proof}
Fix $2 \leq s \leq \eta n$, let us upperbound the probability that there is a set of $s$ constraints containing at most $(K-\delta)s$ variables. Since there are ${\beta n \choose s}$ such sets, the probability is at most
\begin{align*}
&{\beta n \choose s} \Pr [\text{the first $s$ constraints}  \text{contain at most $(K-\delta)s$ variables}]\\
= &{\beta n \choose s} \sum_{i = 1}^{(K-\delta)s} \Pr [\text{the first $s$ constraints} \text{ contain exactly $i$ variables}] .
\end{align*}
Fix a set $T$ of $i$ variables, let $p(s, i)$ be the number of $s$-tuples $(T_1, T_2, \cdots, T_s)$ where for each $1 \leq j \leq s$, $T_j$ is a set of $K$ variables, such that $\cup_{1 \leq j \leq s} T_j = T$. We have
\begin{align*}
&\Pr [\text{the first $s$ constraints contain exactly $i$ variables}] = {n \choose i} \cdot p(i, s) \bigg/ {n \choose K}^s .
\end{align*}
To upperbound $p(i, s)$, we view the way to enumerating valid $(T_1, T_2, \cdots, T_s)$ as, to choose a multiset of $Ks$ variables (each one from $T$) so that each element in $T$ appears at least once in the multiset, then view each element in the multiset as a distinct element, and distribute these $Ks$ elements to $s$ sets, in a balanced way. Note that in this way, we are able to enumerate all the valid $s$-tuples (although some of them might be enumerated more than once). Since there are at most ${Ks -1 \choose i - 1} < {Ks \choose i}$ valid multisets, we have
\begin{align*}
p(i, s) \leq {Ks \choose i} (Ks)!/(K!)^s .
\end{align*}
Therefore, we have
\begin{align*}
&{\beta n \choose s} \Pr [\text{the first $s$ constraints}  \text{contain at most $(K-\delta)s$ variables}]\\
=& {\beta n \choose s}(Ks)!\cdot (K!)^{-s} {n \choose K}^{-s}   \sum_{i = 1}^{(K-\delta)s} {n \choose i} {Ks \choose i} ,
\end{align*}
Note that when $K^2 s < \delta n$ and $i \leq (K -\delta) s$, we have $i < nKs/(n + Ks)$ (since $i \leq Ks(1 - \delta/ K) \leq Ks/(1 + \delta/K) = nKs/(n + \delta n / K)< nKs/(n + Ks)$), and therefore
\begin{align*}
&\frac{{n \choose i}{Ks \choose i}}{{n \choose i - 1}{Ks \choose i - 1}} = \frac{(n-i)(Ks - i)}{i^2} > 1\qquad   (\Leftarrow (n - i)(Ks - i) > i^2 \Leftarrow nKs > (n + Ks) i),
\end{align*}
therefore the function ${n \choose i}{Ks \choose i}$ is increasing when $i \leq (K - \delta) s$, therefore
\begin{align*}
 &{\beta n \choose s}(Ks)!\cdot (K!)^{-s} {n \choose K}^{-s}   \sum_{i = 1}^{(K-\delta)s} {n \choose i} {Ks \choose i} \\
 \leq & {\beta n \choose s}(Ks)!\cdot (K!)^{-s} {n \choose K}^{-s}    \cdot Ks \cdot {n \choose (K-\delta) s} {Ks \choose (K -\delta) s} \\
 = & {\beta n \choose s}(Ks)!\cdot (K!)^{-s} {n \choose K}^{-s}    \cdot Ks \cdot {n \choose (K-\delta) s} {Ks \choose \delta s} .
\end{align*}
for $K \leq n^{1/2}$, we use the fact that ${n \choose K} \geq (n - K)^K/K! \geq n^K/3/((K/e)^K \cdot (5\sqrt{K})) = (en/K)^K / (15\sqrt{K})$ (since by Stirling's formula, we have $K! \leq 5\sqrt{K} (K/e)^K$), and again use the fact that $\sqrt{2 \pi K} (K/e)^K \leq K! \leq 5 \sqrt{K} (K/e)^K$, we bound the expression above by
\begin{align*}
 & \left(\frac{e \beta n}{s}\right)^s \frac{5\sqrt{Ks} (Ks/e)^{Ks}}{(\sqrt{2 \pi K} (K/e)^K)^s} \cdot \left( 15\sqrt{K}  \left(\frac{K}{en}\right)^K  \right)^s  \cdot Ks \cdot  \left(\frac{en}{(K - \delta)s}\right)^{(K-\delta)s} \left( \frac{eK}{\delta}\right)^{\delta s}\\
 \leq& 5 (Ks)^{1.5} \cdot \left( \frac{15 e \beta s^{\delta - 1} K^{K + \delta}}{\sqrt{2\pi } n^{\delta - 1} (K - \delta)^{K - \delta} \delta^\delta }\right)^s\\
 \leq& 5 (Ks)^{1.5} \cdot \left( \frac{15 e^{1 + \delta} \beta s^{\delta - 1} K^{2\delta}}{\sqrt{2 \pi} n^{\delta - 1}  \delta^\delta }\right)^s\\
\end{align*}

For $2 \leq s \leq \ln^2 n$, since $n^{\kappa-1} \leq 1/(10^8 \cdot (\beta K^{2\delta + 0.75})^{1/(\delta - 1)})$, we have $\beta^2 K^{4 \delta + 1.5} / n^{2(\delta - 1)} \leq  n^{-(2\delta - 1)\kappa}$, we have
\begin{align*}
&5 (Ks)^{1.5} \cdot \left( \frac{15 e^{1 + \delta} \beta s^{\delta - 1} K^{2\delta}}{\sqrt{2 \pi} n^{\delta - 1}  \delta^\delta }\right)^s\\
\leq & 5 (Ks)^{1.5} \cdot \left( \frac{15 e^{1 + \delta} \beta s^{\delta - 1} K^{2\delta}}{\sqrt{2 \pi} n^{\delta - 1}  \delta^\delta }\right)^2  \\
\leq & \frac{5 \cdot 15 e^{1 +\delta} s^{\delta - 1}}{\sqrt{2\pi} \delta^\delta n^{2(\delta - 1) \kappa}} \leq  O(n^{-(\delta - 1)\kappa}) .
\end{align*}
For $\ln^2 n < s \leq \eta n$, since $\eta \leq 1/(10^8 \cdot (\beta K^{2\delta + 0.75})^{1/(\delta - 1)})$, we get $\eta \leq 1/(10^8 \cdot (\beta K^{2\delta})^{1/(\delta - 1)})$, and further we have $ \beta K ^{2\delta} \eta^{\delta - 1} \leq \delta^\delta / (100 \cdot 15 e^{1+\delta} / \sqrt{2\pi})$  for all $\delta > 5/4$. Therefore,
\begin{align*}
&5 (Ks)^{1.5} \cdot \left( \frac{15 e^{1 + \delta} \beta s^{\delta - 1} K^{2\delta}}{\sqrt{2 \pi} n^{\delta - 1}  \delta^\delta }\right)^s \\
\leq & 5(Ks)^{1.5}  \left(\frac{s^{\delta - 1}}{100(\eta n)^{\delta - 1}}\right)^s \\
\leq & 5 \cdot \left(\frac{s^{\delta - 1} (Ks)^{(1.5/\ln^2 n)}}{100(\eta n)^{\delta - 1}}\right)^s\\
\leq & 5 \cdot \left(\frac{2}{100}\right)^s. \qquad \qquad \qquad (\text{by $s \leq \eta n$ and $Ks \leq n^2$})
\end{align*}

 Now, we upperbound probability that there exists a set of constraints of size $s \leq \eta n$ involving at most $(K - \delta)s$ variables by
\begin{align*}
& \sum_{s = 2}^{\eta n} 5(Ks)^{1.5} \cdot \left( \frac{15 e^{1 + \delta} \beta s^{\delta - 1} K^{2\delta + 0.5}}{n^{\delta - 1}  \delta^\delta }\right)^s\\
 = &  \sum_{s = 2}^{\ln^2 n} 5(Ks)^{1.5} \cdot \left( \frac{15 e^{1 + \delta} \beta s^{\delta - 1} K^{2\delta + 0.5}}{n^{\delta - 1}  \delta^\delta }\right)^s  + \sum_{s = \ln^2 n + 1}^{\eta n} 5(Ks)^{1.5} \cdot \left( \frac{15 e^{1 + \delta} \beta s^{\delta - 1} K^{2\delta + 0.5}}{n^{\delta - 1}  \delta^\delta }\right)^s\\
\leq& \sum_{s = 2}^{\ln^2 n} O(n^{-\kappa/2}) + \sum_{s = \ln^2n + 1}^{\eta n} 5 \cdot (1 / 50)^s = o(1).
\end{align*} 
\end{proof}

%%%%%%%%%%%%%%%%%%%%%%%%%%%%%%%
\section{Conclusion} \label{sec:conclusion}

In this paper, we show integrality gap lower bounds of $\Omega(n^{1/4}/\log^3 n)$ for $\Omega(\log n/\log\log n)$ levels of the Sherali-Adams+ SDP relaxation, and $\Omega(n^{2/53-\epsilon})$ for $n^{\Omega(\epsilon)}$ levels of the Lasserre SDP relaxation for the \dks~ problem.

The gap instances for SA+ SDP are actually (Erd\"os-Renyi) random graph instances $\mathcal{G}(n, p)$. We believe these instances should give $\Omega(n^{1/4-\epsilon})$ gaps for even stronger relaxations -- in particular higher levels of the Sherali-Adams hierarchy, with stronger SDP constraints. The sub-exponential time algorithms for \dks~ in \cite{BCCFV} imply that the integrality gap becomes $O(n^{1/4 -\eps})$ after $n^{O(\eps)}$ levels of an LP hierarchy which is weaker than the Sherali-Adams hierarchy. In fact, these sub-exponential time algorithms were inspired by attempts to construct integrality gap lower bounds for many levels (polynomial in $n$). It would be interesting to close this gap by obtaining matching integrality gap lower bounds for $n^{\Omega(\eps)}$ levels. As a further goal, one might also hope to combine the techniques used in both parts of this paper, to get $\Omega(n^{1/4-\epsilon})$ gaps for polynomial levels of the Lasserre hierarchy. 

\bibliographystyle{alpha}
%\bibliography{DkS}
\newcommand{\etalchar}[1]{$^{#1}$}

\ifnum\full=1
\begin{appendix}

%\section{Proof of \pref{lem:expansion}}\label{app:expansionproof}

%\section{Proof of Theorem~\ref{thm:lasserre_final}}\label{app:params}

\section{Random graph properties}\label{sec:proofs}
We prove that the properties used in our gap construction hold for $\Gnp$, with $p = n^{-1/2} (\log n)^{1/2}$.  These properties are listed in \pref{sec:instance}.  In what follows fix $p$ to be the value above.  As mentioned in \pref{sec:notation}, the phrase ``with high probability'' (w.h.p.) refers to `with probability at least $1-\frac{1}{q(n)}$', where $q(n)$ is an arbitrary polynomial in $n$ (sometimes there will be a constant depending on the polynomial).

\begin{lemma}
Every vertex of $G$ has degree between $(n^{1/2} \log n)/2$ and $2 n^{1/2} \log n$ w.h.p.
\end{lemma}
\begin{proof}
Let $u \in V$.  The degree $d(u)$ (as a random variable) is the sum of $n$ i.i.d. Bernoulli random variables each having parameter $p=n^{-1/2} \log n$.  The expected value is thus $n^{1/2}\log n$.  This is $\gg \log n$, and thus by Chernoff bounds, the probability that $\Pr[|d(u)- n^{1/2}\log n| > t] \le e^{-t^2/4np(1-p)} < \frac{1}{nq(n)}$, for any polynomial $q(n)$.  Taking union bound gives the claim. 
\end{proof}

\begin{lemma}
Every pair of vertices in $G$ have at most $2\log ^2 n$ common neighbours w.h.p.
\end{lemma}
\begin{proof}
Let $u, v \in V$.  Let $X_i$ be a random variable which is an indicator for $i \in \Gamma(u) \cap \Gamma(v)$.  In $\Gnp$, we have $\E{X_i} = p^2  = \log^2 n/n$.  Thus $\E{ |\Gamma(u) \cap \Gamma(v)|} = np^2 = \log^2 n$.  Thus the probability that it is $>2 \log^2 n$ is at most $e^{-\log^2 n/4}$.  Taking union bound over all $u,v $, we obtain that this is smaller than any polynomial. 
\end{proof}

\begin{lemma}
Every pair of vertices have at least one common neighbour w.h.p.
\end{lemma}
\begin{proof}
As above, consider some $u,v$; we have $\E{ |\Gamma(u) \cap \Gamma(v)|} = np^2 = \log^2 n$.  Thus $\Pr[| \Gamma_u \cap \Gamma(v)| < \log n] \le e^{-\log^2 n/4}$ (since we can use Chernoff bounds as long as the expectation $\gg \log n$).  Taking union bound again implies the result. 
\end{proof}

\begin{lemma}\label{lem:ksoundness}
No induced subgraph on $n^{1/2}$ vertices has density $> 5 \log n$ w.h.p.
\end{lemma}
\begin{proof}
Let $S \subseteq V$ of size $n^{1/2}$.  Then $\E{ E(S, S) } = \binom{n^{1/2}}{2} \cdot p = n^{1/2} \log n/2$.  Further the variance of this quantity is $\binom{n^{1/2}}{2} p(1-p) < n^{1/2} \log n$.  Thus by Chernoff bound,
\[ \Pr[ |E(S,S) - \E{E(S,S)}| >  t] \le e^{-t^2 / 4 n^{1/2} \log n}. \]
Picking $t = 4n^{1/2} \log n$, the probability upper bound is $e^{-4 n^{1/2} \log n}$.  Thus we can take a union bound over all the $\binom{n}{n^{1/2}}$ subsets $S$.  This proves the claim. 
\end{proof}
\end{appendix}
\fi
\end{document}